\documentclass[structabstract]{aa}  
\usepackage{footnote}
\usepackage{rotating}
\usepackage{graphicx}
\usepackage{longtable}
\usepackage[authoryear]{natbib}
\bibpunct{(}{)}{;}{a}{}{,}
\usepackage{txfonts}
\begin{document}
\title{A massive association around the obscured open cluster RSGC3\fnmsep\thanks{Partially based on observations 
collected at the 3.5-m telescope (Calar Alto, Spain)}}
\author{
I.~Negueruela\inst{1}
\and C.~Gonz\'alez-Fern\'andez\inst{1}
\and A.~Marco\inst{1}
\and J. S. Clark\inst{2}
}

\institute{
Departamento de F\'{i}sica, Ingenier\'{i}a de Sistemas y
  Teor\'{i}a de la Se\~{n}al, Universidad de Alicante, Apdo. 99, E03080
  Alicante, Spain\\
\email{ignacio.negueruela@ua.es}
\and
Department of Physics and Astronomy, The Open 
University, Walton Hall, Milton Keynes, MK7 6AA, UK}

\abstract{Four clusters of red supergiants have been discovered in a region of the Milky Way close to base of the Scutum-Crux Arm and the tip of the Long Bar. Population synthesis models indicate that they must be very massive to harbour so many supergiants. If the clusters are physically connected, this Scutum Complex would be the largest and most massive star-forming region ever identified in the Milky Way.}
{The spatial extent of one of these clusters, RSGC3, has not been investigated.  In this paper we explore the possibility that a population of red supergiants could be 
located in its vicinity.}
{We utilised 2MASS $JHK_{{\rm S}}$ photometry to identify candidate obscured luminous red stars in the vicinity of RSGC3. We observed a sample of candidates with the TWIN 
spectrograph on the 3.5-m telescope at Calar Alto, obtaining intermediate-resolution spectroscopy in the 8000--9000\AA\ range. We re-evaluated a number of classification 
criteria proposed in the literature for this spectral range and found that we could use our spectra to derive spectral types and luminosity classes.}   
{We measured the radial velocity of five members of RSGC3, finding velocities similar to the average for members of Stephenson~2. Among the candidates observed outside the 
cluster, our spectra revealed eight M-type supergiants at distances $<18\arcmin$ from the centre of RSGC3, distributed in two clumps. The southern clump is most likely another 
cluster of red supergiants, with reddening and age identical to RSGC3. From 2MASS photometry, we identified four likely supergiant members of the cluster in addition to the five spectroscopically observed. The northern clump may be a small cluster with similar parameters. Photometric analysis of the area around RSGC3 suggests the presence of a large ($>30$) population of red supergiants with similar colours.}
{Our data suggest that the massive cluster RSGC3 is surrounded by an extended association, which may be very massive ($\ga10^{5}\:M_{\sun}$). We also show that supergiants in the Scutum Complex may be characterised via a combination of 2MASS photometry and intermediate-to-high-resolution spectroscopy in the $Z$ band.} 

\maketitle

\section{Introduction}

\defcitealias{davies07}{D07}

After completion of hydrogen core burning, massive stars evolve quickly towards the cool side of the HR diagram,  where they appear as red supergiants (RSGs). Evolutionary 
models show that this transition happens at approximately constant bolometric luminosity, $M_{{\rm bol}}$, resulting  in extremely high infrared luminosities 
\citep{meynet00,marigo}. As a consequence, RSGs are powerful beacons that can be identified even behind huge amounts of obscuration.  Unfortunately, the physical properties of 
RSGs are still poorly constrained \citep{levesque}, although  they are known to span a wide range of luminosities, depending on their initial masses and evolutionary 
status, displaying   
$\log(L_{{\rm bol}}/L_{\odot})\sim4.0$--5.8 \citep{meynet00}. Their infrared spectra, however, do not reflect this variety. All supergiants with spectral types between mid G and late M display very similar infrared spectra, with a strong degeneracy between temperature and luminosity. As a consequence, absolute magnitudes cannot be inferred from infrared spectra, implying that it is not possible to assign physically meaningful parameters to any individual obscured RSG.

In the past few years, a flurry of discoveries has revealed several clusters of RSGs located in a small region of the Galactic plane,
 between $l=24\degr$ and $l=29\degr$ \citep[e.g.,][]{figer06, davies07,
   clark09}. These clusters are very heavily obscured, meaning that their heretofore observed population consists solely of RSGs. However, population synthesis models suggest that the clusters must be very massive to harbour these populations \citep[e.g.,][]{davies07,simonw51}. The first cluster found, RSGC1 \citep{figer06}, is the most heavily obscured, and also the youngest at an estimated $\tau$=12$\pm2\:$Myr and $M_{{\rm initial}}$=3$\pm1\times10^{4}M_{\sun}$ \citep{davies08}. Alicante~8, located in its immediate vicinity, seems slightly older and rather smaller \citep{neg10}. RSGC2 = Stephenson~2 (Ste~2) is the least obscured and
apparently most massive of all, with $\tau$=17$\pm3$~Myr and $M_{{\rm
    initial}}$=4$\pm1\times10^4M_{\odot}$ (\citealt{davies07}; henceforth \citetalias{davies07}).  Finally,
RSGC3 lies at some distance from all the others and has an estimated
$\tau$=16--20$\,$Myr and an inferred $M_{{\rm initial}}$=2--4$\times10^4M_{\sun}$  \citep{clark09,alexander09}. Collectively,
 these clusters host a significant fraction of the known RSG population in the Galaxy.

The line of sight to the RSG clusters (RSGCs) passes through  several Galactic arms and reaches an area of very high obscuration at a distance comparable to those of the 
clusters \citep{neg10}. Stellar densities are extremely high  at moderate $K$ magnitudes, meaning that the cluster populations are very hard to identify, except for the 
RSGs. Because of this, membership of individual stars is difficult  to assess and the actual extent of the clusters is uncertain. Membership in RSGC1 and Ste~2 has been 
defined in terms of a common radial velocity, $v_{{\rm rad}}$, measured from high-resolution $K$-band spectra \citep{davies07,davies08}. In the case of RSGC1, the 
obscuration 
is very high and the RSG members are intrinsically very bright, resulting in a relatively easy selection of candidate members. On the other hand, the line of sight to 
Ste~2 is very rich in bright red stars. \citetalias{davies07} find a large number  of foreground giants and also several putative RSGs with $v_{{\rm rad}}$ 
apparently excluding membership, 
even though their analysis was confined to a circle with radius $r=7\arcmin$ around the nominal cluster centre. For the other clusters, dynamical analysis has not yet been 
carried out and the determination of membership is less secure.

With the data currently available, the exact nature of this Scutum Complex is unclear. There have been suggestions that it represents a giant star formation region triggered by the dynamical excitation of the Galactic Bar, whose tip is believed to intersect the Scutum-Crux Arm close to this region
(see discussion in \citetalias{davies07}; \citealt{garzon}). If all the RSG clusters are located at a common nominal distance of 6.5~kpc, they span $\sim 500$~pc in projection. Alternatively, if there is a  giant Molecular Ring at the end of the Bar \citep[e.g.,][]{rathborne09}, our sightline could cut its cross section over a distance $\sim3$~kpc. 

Heretofore, all studies of the RSGCs have been carried out solely with infrared data. Intermediate-resolution $K$-band spectroscopy has been used to observe the CO bandhead  absorption features near 2.3$\mu$m, which identify cool, evolved stars. The equivalent width (EW) of the first bandhead feature is a valid estimator of spectral type at intermediate resolution, though the relationship is bimodal (showing a similar slope for giants and supergiants, but higher EWs for the latter), and scatter is large \citep{davies07,neg10}. \citet{alexander09} explore a similar relation for low-resolution spectra. In both cases, the accuracy in the spectral-type determination is estimated at $\pm2$ subtypes. Until the advent of infrared spectrographs with multiplexing capabilities, extending this methodology to large samples will consume large amounts of telescope time. 

Optical spectrographs with high multiplexing factors do, however, exist. Highly obscured  RSGs are, unfortunately, too faint to be detected in the optical range. The 
far  red optical region, between 8000\AA\ and 10000\AA\ offers a good compromise, as it can be reached with optical instrumentation and has lower extinction. By using 
existing 
catalogues, namely B/DENIS and USNO-B1.0 \citep{monet03}, we have verified that more than 50\% of the confirmed RSG members of Ste~2 and RSGC3 are brighter than $i=16$ and 
can be observed at intermediate-high resolution with existing instrumentation.

In this paper, we present a pilot study to address several issues that will  allow us to evaluate the possibility of detecting RSGs over a wide area by using optical 
spectrographs. Firstly, we explore the use of photometric criteria and catalogue cross correlation to detect candidate luminous red stars. Secondly, we re-evaluate several 
criteria 
presented in the literature to estimate spectral types and luminosities with intermediate-resolution far red optical  spectra. Finally, we assess the existence of an 
extended population 
of RSGs around RSGC3 that may justify wide-field searches. The paper is organised as follows. In Section~\ref{sec:target}, we set photometric criteria for selection of candidates and describe the observations of a sample of candidates. In Section~\ref{sec:calib}, we discuss several methods of classification for red stars in the near-IR. In Section~\ref{sec:res}, we present the analysis of our observations. Finally, we discuss their implications in Section~\ref{sec:discu}.

\section{Target selection and observations}
\label{sec:target}

RSGC3 was discovered as a concentration of very bright infrared sources \citep{clark09,alexander09}.
\citet{clark09} explored a circle of radius $r=7\arcmin$ around the nominal centre of RSGC3, finding a strong concentration of RSGs towards the centre, but also several photometric candidate members at the edges of their search area. \citet{alexander09} analysed a smaller area.
 
In regions of high extinction, traditional photometric criteria cannot always  be employed  to select stars of a particular spectral type and luminosity class. While RSGs are very bright in $K$, several classes 
of red giants are also bright infrared sources, but much more numerous. Despite this, if we are to detect RSGs in the area surrounding RSGC3, we might expect them to have similar photometric characteristics to known cluster members.

\subsection{Photometric criteria}
\label{sec:2mass}

 We have used 2MASS $JHK_{{\rm S}}$ photometry \citep{skru06} to identify possible RSG stars. As discussed in \citet{neg10}, the reddening-free parameter
$Q=(J-H)-1.8\times(H-K_{{\rm S}})$ allows the separation of early-type stars ($Q\la0.0$) and late-type main-sequence stars and  red  
clump giants ($Q\ga 0.4$). RSGs show a peculiar behaviour and present very different $Q$ values. Most of them deredden to the position of yellow stars, with $Q\approx0.2-0.3$, though a fraction fall close to the position of red stars ($Q\approx0.4$). 

For this pilot programme, we  selected stars within $20\arcmin$ of the nominal centre of RSGC3, with $(J-K_{{\rm S}})>1.3$. The colour cut was chosen to be redder than the intrinsic colours of any normal, unreddened star, as we expected our objects to suffer significant extinction. On the other hand, we did not want to force candidates to have the same colours as cluster 
members, since extinction might be variable on small scales over this field. We selected stars brighter than $K_{{\rm S}}=7.0$, as RSGs are expected to be bright. Typical 
magnitudes for confirmed members of RSGC3 lie in the $K_{{\rm S}}=5-6$ range, and so $K_{{\rm S}}=7.0$ allows for much heavier extinction. Finally, we imposed $0.1\leq Q 
\leq 0.4$, the range where most RSGs lie, in order to exclude foreground red-clump giants and main-sequence M-type stars. We note that the criteria are unlikely  to select 
{\it all} RSGs, as some of them have colours $Q\approx0.4$, or may be so heavily reddened as to have $K_{{\rm S}}>7.0$. The criteria were chosen to select most RSGs, while 
leaving out many likely interlopers. 

\begin{figure*}
\resizebox{\textwidth}{!}{
\includegraphics[clip]{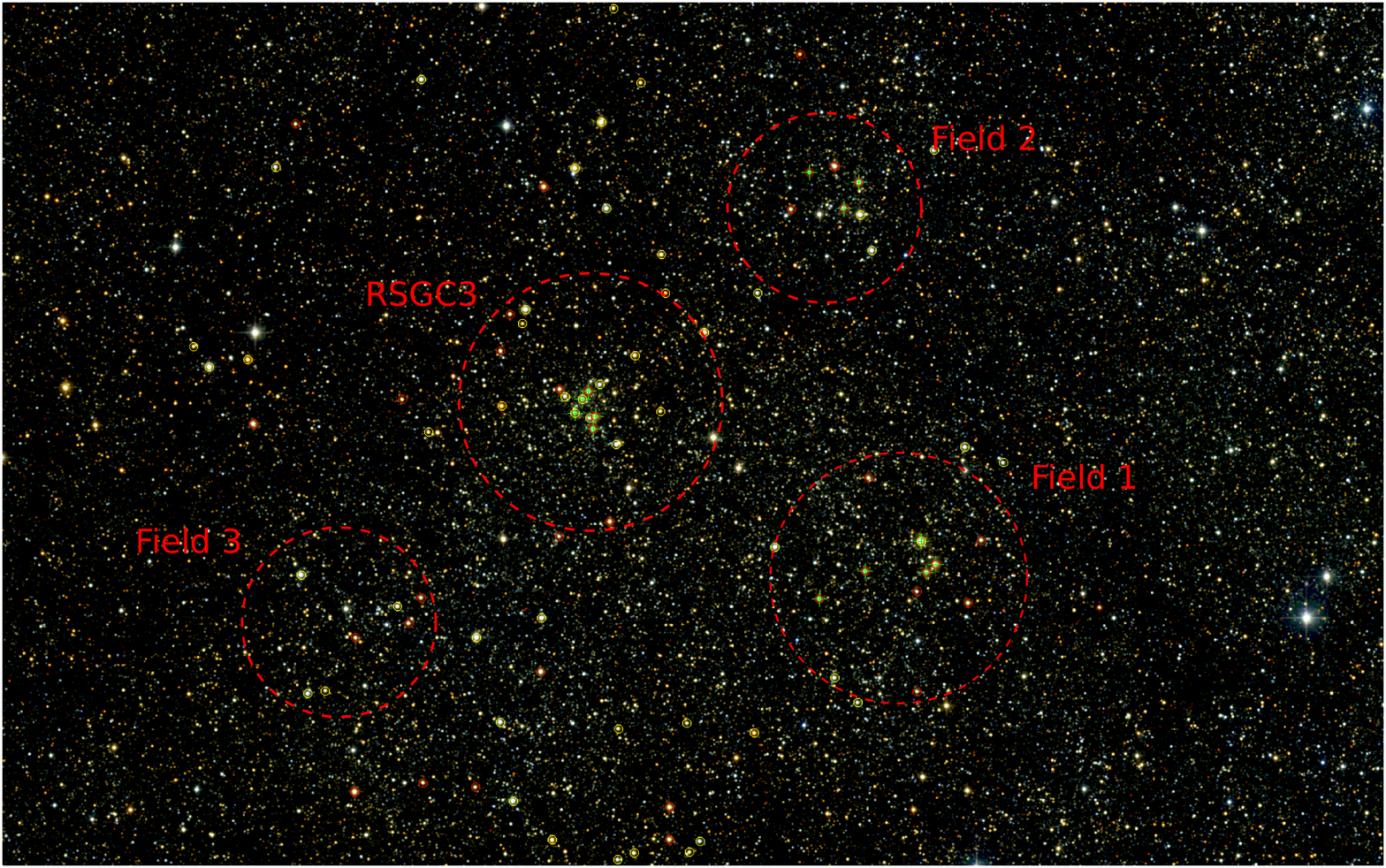}}
\caption{Colour-composite 2MASS $JHK_{{\rm S}}$ finder of the area surveyed. The image covers $\sim55\arcmin\times35\arcmin$ (roughly corresponding to $100\times60$~pc at 6~kpc).  The circle showing the position of RSGC3 has radius $6\arcmin$. Yellow circles show stars that pass all the photometric criteria in Section~\ref{sec:2mass}. Red circles indicate stars which, in addition, are detected as DENIS or USNO.B-1 sources with $i>13.2$ (see Section~\ref{extended} for this selection criterion). Crosses indicate the stars spectroscopically identified as reddened RSGs in this paper. North is up and East is left.\label{finder}}
\end{figure*}

We then checked our selection against the B/DENIS and USNO-B1.0 catalogues. Stars clearly detected in the DSS2 blue plates were eliminated from the sample. In addition, we 
set a cut at $i>9.5$, as we did not expect our targets to be brighter than this limit, given that extinction is much higher in $i$ than in $K$. The size of the telescope 
forces us to a limit $i<15.5$. This limit was never reached during the observations, as the seeing was relatively poor throughout the run. Only stars brighter than $i=14.5$ were observed.

\begin{table*}
\caption{Candidates selected as possible red luminous stars, with their characteristics.\label{tab:fields}}
\begin{tabular}{rcccccccccc}
\hline
\hline
ID & 2MASS & $J$\tablefootmark{1} & $H$\tablefootmark{1}    & $K_{{\rm S}}$\tablefootmark{1} &  $Q$ & $i$\tablefootmark{1}& Spectral &Spectral & $E(J-K_{{\rm S}})$ &$v_{{\rm LSR}}$\\
& & & & & & &Type (TiO) & Type & & (km\,s$^{-1}$)\\
\hline
\noalign{\smallskip}
\multicolumn{11}{c}{Field 1}\\
\noalign{\smallskip}
\hline
F1S01  & 18445403$-$0329009 & 7.55$\pm$0.02 & 6.26$\pm$0.04 & 5.71$\pm$0.02 & 0.30 & 11.08$\pm$0.02 &M7 & M7\,Ib/II& 0.55 &+43\\
F1S02  & 18444686$-$0331074 & 8.78$\pm$0.03 & 6.71$\pm$0.02 & 5.76$\pm$0.02 & 0.38& 14.44$\pm$0.04 &M1 & M1\,Ia$^{*}$&1.96& +73\\
F1S03  & 18444439$-$0334197 & 8.21$\pm$0.02 & 6.86$\pm$0.04 & 6.32$\pm$0.02 & 0.39& 11.30$\pm$0.02&$-$&$-$&$-$&$-$\\
F1S04  & 18443941$-$0330003 & 8.66$\pm$0.03 & 6.76$\pm$0.04 & 5.82$\pm$0.02 & 0.19& 13.99$\pm$ 0.03&M2 &M2\,Iab$^{*}$&1.74& +112\\
F1S05  & 18443100$-$0330499 & 9.87$\pm$0.02 & 7.93$\pm$0.05 & 6.97$\pm$0.03 & 0.21& 15.44$\pm$0.06&$-$&$-$&$-$&$-$\\
F1S06  & 18442945$-$0330024 & 8.45$\pm$0.03 & 6.39$\pm$0.03 & 5.41$\pm$0.02 & 0.30& 14.13$\pm$0.04&M3 &M2\,Ia$^{*}$&1.94& +93\\
F1S07  & 18442796$-$0329425 & 8.06$\pm$0.02 & 5.91$\pm$0.04 & 4.85$\pm$0.02 & 0.25& 14.46$\pm$0.04&M5 &M4\,Iab$^{*}$&2.04&+93\\
F1S08  & 18443037$-$0328470 & 7.34$\pm$0.02 & 5.20$\pm$0.04 & 4.10$\pm$0.30 & 0.16 & 13.70$\pm$0.03 &M4&M4\,Ib$^{*}$&2.07&+75\\
F1S09  & 18442053$-$0328446 & 9.28$\pm$0.02 & 7.24$\pm$0.07 & 6.21$\pm$0.03 & 0.21& 15.32$\pm$0.06 &$-$&$-$&$-$&$-$\\
F1S10 & 18442322$-$0324578 & 8.08$\pm$0.03 & 6.64$\pm$0.05 & 5.94$\pm$0.02 & 0.18& 11.91$\pm$0.02&M6 &M5\,Ib/II& 0.95&+59\\
F1S11 & 18441700$-$0325360 & 8.68$\pm$0.03 & 7.40$\pm$0.04 & 6.89$\pm$0.02 & 0.36& 11.44$\pm$0.02&$-$&$-$&$-$&$-$\\
F1S12 & 18444065$-$0335199 & 8.90$\pm$0.03 & 7.49$\pm$0.03 & 6.90$\pm$0.02 & 0.37& 12.22$\pm$0.03&$-$&$-$&$-$&$-$\\
\hline
\noalign{\smallskip}
\multicolumn{11}{c}{Field 2}\\
\noalign{\smallskip}
\hline
F2S01  & 18444846$-$0313495  & 9.14$\pm$0.02  & 7.33$\pm$0.03  & 6.50$\pm$0.02&  0.32 &14.01$\pm$0.04&M1 & M0\,Ia$^{*}$&1.63 & +72\\
F2S02  & 18444431$-$0313350  & 8.69$\pm$0.02  & 6.39$\pm$0.03  & 5.23$\pm$0.02  &0.22& 15.52$\pm$0.07&$-$&$-$&$-$&$-$\\
F2S03  & 18444044$-$0314142  & 8.06$\pm$0.02  & 6.22$\pm$0.03  & 5.31$\pm$0.02   & 0.21&13.21$\pm$0.03&M1 &M1\,Iab$^{*}$&1.67& +86\\
F2S04  & 18444283$-$0315168  & 8.53$\pm$0.02  & 6.69$\pm$0.04  & 5.80$\pm$0.02  & 0.24&13.69$\pm$0.03&M2 &M1.5\,Iab$^{*}$&1.64& +87\\
F2S05  & 18444023$-$0315329  & 7.21$\pm$0.03  & 5.51$\pm$0.03  & 4.72$\pm$0.03  & 0.30&12.40$\pm$0.03&M8 &M7\,II& 1.19& +61\\
F2S06  & 18442823$-$0312558  & 7.50$\pm$0.03  & 6.38$\pm$0.03  & 5.94$\pm$0.02 & 0.34&9.70$\pm$0.03&$-$&$-$&$-$&$-$\\
F2S07  & 18443833$-$0316596  & 7.32$\pm$0.02  & 6.05$\pm$0.04  & 5.49$\pm$0.02&  0.26& 10.68$\pm$0.03&M7& M6\,III&0.58& +85\\
F2S08  & 18443324$-$0306334  & 8.44$\pm$0.03  & 7.34$\pm$0.06  & 6.88$\pm$0.04  & 0.27&10.73$\pm$0.02&$-$&$-$&$-$&$-$\\
F2S09  & 18443220$-$0304597  & 9.04$\pm$0.03  & 7.29$\pm$0.05  & 6.44$\pm$0.04  & 0.23&13.77$\pm$0.03&M5 &M5.5\,III&1.36& +94\\
F2S10 & 18443172$-$0302217  & 8.40$\pm$0.02 & 6.85$\pm$0.03 & 6.19$\pm$0.03& 0.35&12.15$\pm$0.03&$-$&$-$&$-$&$-$\\
F2S11 & 18444658$-$0302416  & 8.65$\pm$0.02  & 6.69$\pm$0.03 & 5.79$\pm$0.02   &0.33& 14.18$\pm$0.04&$-$&$-$&$-$&$-$\\
\hline
\noalign{\smallskip}
\multicolumn{11}{c}{Field 3}\\
\noalign{\smallskip}
\hline
F3S01&18461105$-$0330091 & 7.08$\pm$0.02  & 5.79$\pm$0.03  & 5.26 $\pm$0.02  & 0.33 & 10.04$\pm$0.04&M3 &M4\,II-III& 0.65 &$-$22\\
F3S02&18460994$-$0334581 & 8.10$\pm$0.02  & 6.80$\pm$0.05  & 6.26$\pm$0.02  & 0.33 & 10.94$\pm$0.02&M2 &M3.5\,III& 0.70 & +10\\
F3S03&18455541$-$0331254 & 8.37$\pm$0.02  & 6.82$\pm$0.05  & 6.17 $\pm$0.02  & 0.38 & 13.04$\pm$0.02&M8 &M7\,III & 0.90& +66\\
F3S04&18455357$-$0332069 & 9.30$\pm$0.02  & 7.31$\pm$0.04  & 6.38$\pm$0.03  & 0.31 & 14.76$\pm$0.04&$-$&$-$&$-$&$-$\\
F3S05&18454257$-$0332406 & 6.68$\pm$0.02  & 5.27$\pm$0.02  & 4.60$\pm$0.02  & 0.19 & 11.67$\pm$0.02&M9 & M7\,III& 0.78& $-$37\\
F3S06&18453198$-$0331536 & 7.99$\pm$0.02  & 6.31$\pm$0.03  & 5.51$\pm$0.03  & 0.24 & 11.5$\pm$0.3&$-$&$-$&$-$&$-$\\
F3S07&18453212$-$0334044 & 8.40$\pm$0.03  & 6.45$\pm$0.03  & 5.51$\pm$0.03  & 0.25 & 13.4$\pm$0.3&$-$&$-$&$-$&$-$\\
F3S08&18453872$-$0336069 & 7.55$\pm$0.03  & 6.23$\pm$0.033  & 5.66$\pm$0.02   & 0.29 & 10.96$\pm$0.02&M6 &M5\,II-III&0.68&+59\\
F3S09&18453660$-$0339191 & 7.20$\pm$0.02  & 5.83$\pm$0.04  & 5.24$\pm$0.02  & 0.30 & 10.42$\pm$0.03&M3 &M4\,II&0.79& +35\\
F3S10&18460233$-$0338568 & 8.54$\pm$0.03  & 6.10$\pm$0.04  & 4.94$\pm$0.02  & 0.36 & 15.31$\pm$0.05&$-$&$-$&$-$&$-$\\
 \hline
\end{tabular}
\newline\\
\tablefoottext{1}{$JHK_{{\rm S}}$ magnitudes from 2MASS. $i$ magnitudes from DENIS, except for F3S06 and F3S07, which are photographic $I$ from USNO.}
\newline\\
\tablefoottext{*}{Likely members of the RSGC3 association.}
\end{table*}

There are $\sim50$ stars fulfilling these criteria and lying at distances $>7\arcmin$ and $<20\arcmin$ from the cluster centre. They are not evenly spread, and we can 
identify three areas of high density, which we select for systematic observations (see Fig.~\ref{finder}). Field~1 lies $14\arcmin$ South-West of RSGC3. There is a compact group of candidates with several other objects around. 
Field~2 contains an obvious   concentration  of very bright (in $K$) stars, most of which pass all the photometric criteria. It is located $\sim12\arcmin$ North-West of 
the  centre  of RSGC3. We also include a few fainter stars a few arcmin to the North. Finally, Field~3 does not appear to contain any obvious aggregate, instead
comprising an area of uniformly distributed photometric candidates. It lies $\sim 12\arcmin$ South-East of the cluster core. 
All the 
photometric 
candidates 
are listed in Table~\ref{tab:fields}, where they are given an identifier of the form FnSnn, where the first digit identifies the field and the last two digits identify the individual star.

\subsection{Spectroscopy}

Observations were carried out with the Cassegrain Twin Spectrograph (TWIN) of the 3.5 m telescope at Calar Alto (Almer\'{\i}a, Spain), during a run on 2009 July 10--11. We used the red arm equipped with the SITe\#20b\_12 CCD and the T06 (1200 ln\,mm$^{-1}$) grating. The grating angle was chosen to cover the 8000--9000\,\AA\ range. The $1\farcs2$ slit gives a resolution element of 1.2\AA, measured on the arc lamps. This is equivalent to a resolving power $R=7\,000$.  
For this pilot run, we chose a longslit spectrograph, as it allows a much better selection of individual targets and observation of stars in the crowded cluster cores.

In addition to the stars in the three fields listed above, we observed seven objects from the list of members of Ste~2 in \citetalias{davies07} to be used as references for radial velocity measurements, and five photometric members of RSGC3. 

 The seeing was variable, with typical values $\approx1\farcs5$ (the monitor was not active at the time, so only rough estimates were available). Transparency was high, except for the first two hours of the first night, when some clouds were present close to the horizon. Typical exposure times ranged from 150~s for the brightest objects ($i\la10$) to 2000~s for objects fainter than $i=14.0$. To reduce integration times, most targets were observed in pairs, aligned along the slit. Because of this, parallactic angles were not used. ThAr arc lamps were observed immediately before or after any exposure, and each time that the telescope was moved more than a few arcmin. All exposures have been calibrated with, at least, one arc taken less than 1800~s before or after the midpoint of the observation and, in many cases, are sandwiched between two arcs.

All the data have been reduced using the {\em Starlink}
software packages {\sc ccdpack} \citep{draper} and {\sc figaro}
\citep{shortridge} 
and analysed using {\sc figaro} and {\sc dipso} \citep{howarth}.  Arc frames were extracted at the position of each target star, and the respective solutions checked. Typical RMS values for the fit to a third-degree polynomial are $\approx0.1$\AA. Solutions at different positions within a given frame are significantly shifted, with differences up to two pixels.

\begin{figure}
\resizebox{\columnwidth}{!}{
\includegraphics[angle=-90,clip]{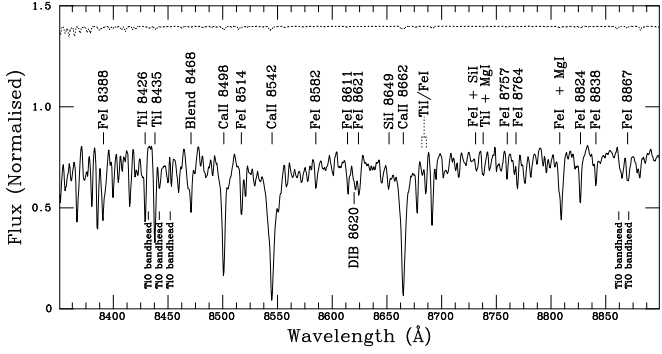}}
\caption{Intermediate resolution spectrum of star D18 in Ste~2, with features identified, and the position of the TiO bandheads used for spectral classification of later-type stars marked. This object is M1.5\,Iab according to the features in this spectral region.  Also shown (displaced upwards by 0.4 continuum units) is the sky transmission in this spectral region, from \citet{hinkle}.\label{muestra}}
\end{figure}

The average signal-to-noise ratio (SNR) of our targets was not high, typically 50--70 per pixel, though higher for the brightest targets, and lower for some objects (a few 
spectra have SNR$\sim35$). As an example, in Fig.~\ref{muestra}, we show the spectrum of star D18 (i.e., star \#18 from \citetalias{davies07}) in Ste~2, which has a SNR $\sim90$ per pixel ($\sim145$ per resolution element), with a number of features identified, after different references in the literature, notably \citet{kirkpatrick91}, \citet{mallik97}, and \citet{cenarro01}.

\section{Spectral classification}
\label{sec:calib}

The classification of red luminous stars in the far red region has been explored by several authors. Different criteria have been proposed depending on the resolution available. For example, \citet{boschi}, using echelle spectra \citep{munarit99,marrese}, propose classification criteria based on individual line ratios. At lower resolution, \citet{ginestet94} and \citet{carquillat97} investigated the variation of different line strengths with temperature and luminosity with the intention of finding valid criteria for the determination of effective temperature ($T_{{\rm eff}}$) and luminosity. 

In general, at a given metallicity, the strength of the metallic  spectrum increases with luminosity \citep{ginestet94,carquillat97}. An important case in point is the 
strength  of the Ca\,{\sc ii} triplet. These lines dominate the $I$-band spectra of stars with spectral types between early F and M3 and their strength is strongly 
correlated with luminosity \citep{diaz89,zhou91}, although they also demonstrate a  dependence on metallicity and spectral type. Several authors have studied the 
dependence of the equivalent 
width (EW) of these lines with luminosity class, but the results depend to some degree on the resolution used and the way in which the continuum is defined \citep[cf.][]{mallik97}.

While main sequence stars are easy to tell apart, differences between giants and supergiants are more subtle. Amongst several line strengths and line ratios studied by 
\citet{carquillat97}, the strength of the  Ti\,{\sc i} and Fe\,{\sc i} blend at 8468\AA\ seems to be the best discriminant at a given spectral type. However, all luminosity criteria show a certain degree of dispersion due to differences in chemical composition. In addition, many metallic lines show dependence on both temperature and luminosity. As a consequence, the boundary between luminosity classes II and Ib may be uncertain \citep{ginestet94,zhu99}, in particular for late K types \citep{ginestet94}. 

The different luminosity calibrations in the literature have been derived from spectra with different resolutions, using diverse measuring methods. Moreover, all these 
works have 
used spectra with high SNR. As our spectra have moderately high resolution and only moderate SNR, we choose to measure the EW of lines relative to the local  continuum, 
 using windows defined by the apparent line width (which, for example, turn out to be $\approx$14\AA\ for Ca\,{\sc ii}~8498\AA\ and $\approx$18\AA\ for Ca\,{\sc 
ii}~8542\AA\ in all our stars).  As there is an amount of subjectivity in this definition, we  estimate  the {\it internal} errors in the measurements by taking a large 
number of measurements with small  differences in the definition of the continuum and integration limits. Typically the errors are found to be $\approx3$\%. To calibrate 
our measurements, we utilised the spectral library of \citet{cenarro01} since the spectra in this library have similar resolution to ours\footnote{The spectra in 
\citet{cenarro01} were observed in different runs at several telescopes. Around two thirds of the spectra have resolution (FWHM) of 1.2\AA, around one third have resolution of 1.5\AA, while a few have lower resolutions.}. In addition, many of the stars included are defined as secondary MK standards in \citet{keenan} or have accurate spectral types in \citet{kp80}. We find 38 K-type stars and 14 M-type stars (earlier than M6) classified as giants or supergiants in these references.

\subsection{Temperature determination and spectral types}
\label{sec:teff}

\begin{figure}
\resizebox{\columnwidth}{!}{\includegraphics[angle=90,clip]{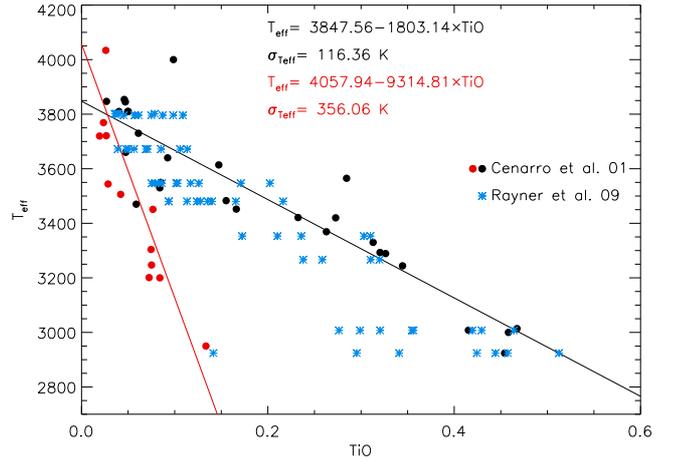}
}
\caption{Calibration of the depth of the TiO bandhead at 8660\AA\ against $T_{{\rm eff}}$ for a sample of dwarfs (red circles), giants and supergiants (black circles) with MK spectral types present in the library of \citet{cenarro01}. Adding stars from the library of \citet{rayner09}, observed at lower resolution (asterisks), does not significantly change the trend.\label{fig:tiobands}}
\end{figure}

For late-type stars the dominant features in the $I$-band spectra are the molecular TiO absorption bands, with VO bands also becoming prominent for spectral types M6 and later \citep{kirkpatrick91}. These bands are much stronger in luminous stars (giants and supergiants) than in stars close to the main sequence. The strength of the TiO bands has been used in the past as the basis of several temperature calibrations, via the definition of different photometric or spectroscopic indices \citep[see for example][]{am97,cenarro09}. \citet{ramsey} found that the depth of the TiO $\delta$(0,0) $R$-branch bandhead at 8660\AA\ correlated strongly with $T_{{\rm eff}}$ for a sample of K and M giants observed at high resolution. Here we build our own temperature scale based on the strength of this feature on intermediate-resolution spectra. We measure the TiO absorption between 8860\AA{} and 8872\AA{} against a continuum estimation between 8850\AA{} and 8857\AA{}. The magnitude of this jump is then calculated as:
\begin{equation}
TiO=1-\frac{I_{{\rm TiO}}}{I_{{\rm cont}}}
\, .
\end{equation}

To calibrate $T_{{\rm eff}}$ against this measurement of the TiO band strength, we used the spectral library of \citet{cenarro01} as a source of primary calibrators. In 
order to check if there is a 
dependence on spectral resolution, we also employed the library of \citet{rayner09}. This catalogue covers our spectral range at $R=2\,000$, and again contains mostly 
stars 
classified in \citet{keenan}. Both catalogues contain a fairly wide sample of spectral types, including giants, supergiants and dwarfs. Over the  spectral range and at
the resolution in question, giants and supergiants have very similar spectra. Therefore we explored the possibility that a single calibration can be used for luminous 
stars of all 
luminosity classes. As \citet{cenarro01} provide $T_{{\rm eff}}$ determinations for all their stars, we used  this subset to obtain both a relation between TiO band depth 
and $T_{{\rm eff}}$ and a relation between $T_{{\rm eff}}$ and $G$, an index that labels the spectral type of  the star ($G=0$ for a putative O0 star, increasing by one 
for each type, reaching $G=30$ for an F0 star and so on). We found that for types later than A0, 
\begin{equation}
\label{caltyp}
\log{(T_{{\rm eff}})}=(4.155\pm0.008)-(9.79\pm0.16)\times \frac{G}{1000}
\, .
\end{equation}

Initially, we fit the relationship between the TiO index and $T_{{\rm eff}}$ for the stars from \citet{cenarro01}.
The results are plotted in Fig.~\ref{fig:tiobands}. As is seen, the index correlates well with temperature for spectral types later than M0; the resultant  relation being
\begin{equation}
\label{calteffg}
T_{{\rm eff}}=(3848\pm38)-(1803\pm154)\times TiO\, .
\end{equation}

Giants and supergiants cannot be separated. 

For stars in the catalogue of \citet{rayner09}, we only have spectral types. Consequently, we first  estimated their $T_{{\rm eff}}$ via Eq.~\ref{caltyp}. With this 
estimate, we then checked that they follow the same ($TiO$,$T_{{\rm eff}}$) relation as stars in \citet{cenarro01} (Fig.~\ref{fig:tiobands}). 
Typical errors are $\pm116\:$K for the stars from \citet{cenarro01} and  $\pm143\:$K for stars in \citet{rayner09}. Both dispersions are fully compatible, as the latter value includes the errors from both Eqs.~\ref{caltyp} and~\ref{calteffg}. Giants and supergiants follow the same relation, while the difference in spectral resolution between the two samples does not seem to affect the index significantly\fnmsep\footnote{In order to explore the effect of spectral resolution, we  artificially degraded the resolution of the spectra in \citet{cenarro01} to lower resolutions. We find that the $TiO$ index starts to loose sensitivity to the earliest M spectral types at resolving power $R\la1500$ and deviates strongly from the  relation shown in Eq.~\ref{calteffg} for $R\la1000$.}. 

 Main-sequence stars, on the other hand, have rather weaker TiO features, and follow a very different relation,
\begin{equation}
\label{calteffd}
T_{{\rm eff}}=(4058\pm140)-(9315\pm2190)\times TiO \, .
\end{equation}

Because of the weakness of the TiO band, the relation shows an increased dispersion, with typical errors $\pm356\:$K for stars from \citet{cenarro01}.

\subsection{Luminosity calibration}
\label{sec:lum}

\begin{figure}
\resizebox{\columnwidth}{!}{
\includegraphics[angle=0,clip]{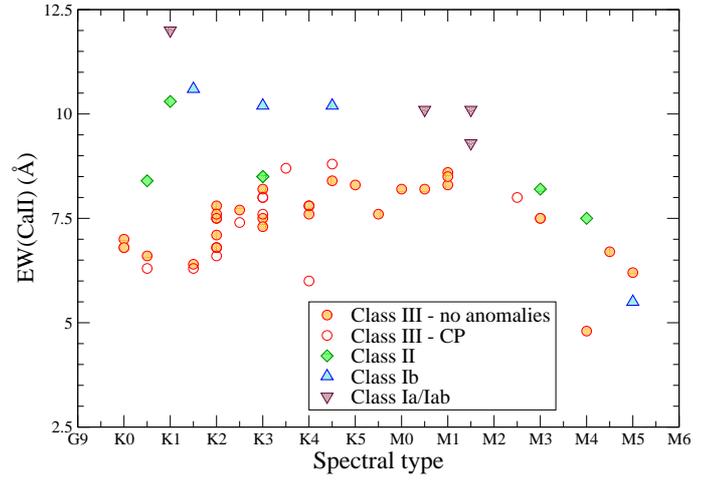}}
\caption{Measurements of the EW of \ion{Ca}{ii} for stars in the library of \citet{cenarro01}. Circles are luminosity class III giants (open circles are stars whose spectral classification includes some chemical anomaly, while filled circles have no such qualifiers). Diamonds are luminosity class II bright giants. Triangles are supergiants (triangles pointing upwards have luminosity class Ib, while those pointing downwards have class Iab). With one single exception, giants and supergiants are well separated up to spectral type M2.  The strongest \ion{Ca}{ii} triplet corresponds to the K1\,Ia--Iab supergiant HD~63302.}\label{caii}
\end{figure}

We then proceeded to calibrate the main luminosity indicators, namely the strength of the Ca\,{\sc ii} triplet and the Ti\,{\sc i} dominated blend at 8468\AA. For this, we 
measured  the EWs of these features on standard  stars in \citet{cenarro01}, using the same measuring technique as for our targets.  As the number of supergiants is 
low in this sample, we complement the dataset with a few more stars from \citet{ginestet94}, at slightly higher resolution. The results for the two strongest \ion{Ca}{ii} lines are displayed in Fig.~\ref{caii}.
We find ``normal'' luminosity class III stars to have EW(Ca\,{\sc ii}~8542+Ca\,{\sc ii}~8662) in the range 6--8\AA, with a clear tendency to lower values at early K types. Stars whose classification includes metallicity anomalies deviate slightly from the general trend. Ba stars generally have high values, but all giants have EW(Ca\,{\sc ii}~8542+Ca\,{\sc ii}~8662)$<$8.5\AA. The few luminosity class II bright giants in the sample have EW(Ca\,{\sc ii}~8542+Ca\,{\sc ii}~8662)$\approx$8.5\AA, while the seven supergiants earlier than M3 have EW(Ca\,{\sc ii}~8542+Ca\,{\sc ii}~8662)$>$9\AA. These results agree qualitatively with those of \citet{ginestet94}, though our definition of the continuum leads to smaller EWs\fnmsep\footnote{\citet{ginestet94} use a few high points in the spectra to define the continuum, while most other references \citep[e.g.,][]{diaz89,mallik97} use small featureless sections at the edges of the spectral range. As a consequence, \citet{ginestet94} measure higher EWs than other authors do.}. The EW increases on average with spectral type, until it reaches a maximum at K4--M1, and then decreases, due to the depression of the continuum level by the TiO features The criterion defined by \citet{diaz89} to separate supergiants from bright giants at solar composition, namely EW(\ion{Ca}{ii}~8542+\ion{Ca}{ii}~8662)$>$9\AA, also appears valid for our resolution and continuum definition, at least for spectral types earlier than M3.

\begin{figure}
\resizebox{\columnwidth}{!}{
\includegraphics[angle=0,clip]{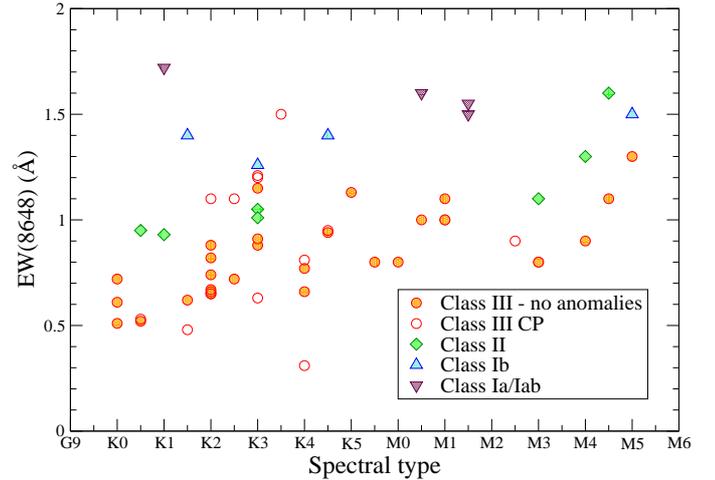}}
\caption{Measurements of the EW of \ion{Ti}{i}/\ion{Fe}{i} blend for stars in the library of \citet{cenarro01}. Symbols are as in Fig.~\ref{caii}. Except for a few deviant points at K3, supergiants are very well separated from normal-composition giants. The strongest feature is measured again in HD~66302.\label{blend}}
\end{figure}

Figure~\ref{blend} shows the measurements of the 8468\AA\ blend. We find that ``normal'' class III stars have EW(8468) in the 0.6--0.8\,\AA\ range, again with the lower values corresponding to the earlier types. Metallicity anomalies cause important dispersion here. In particular, stars classified as CN-strong have values in the 1.0--1.2\,\AA\ range. These objects are believed to be to some degree metal-rich \citep{kh94}.
All bright giants have EW(8468)=0.9--1.1\,\AA, with all supergiants displaying values $>1.2$\AA. Our values agree very well with those of \citet[their Fig.~13]{ginestet94}, as the local continuum in this range coincides with their selected high points. We can thus confirm that the EW of this blend is a very good luminosity indicator, allowing perhaps a better separation of supergiants and giants of normal composition than the \ion{Ca}{ii} lines.
Interestingly, all the CN-strong giants have EW(Ca\,{\sc ii}~8542+Ca\,{\sc ii}~8662)$\approx8$\AA, high for class III, but always lower than supergiants\footnote{This is not surprising, as the high EWs of the Ca\,{\sc ii} lines in supergiants are due to the development of broad wings at low gravity}. Therefore a combination of EW(Ca\,{\sc ii}) and EW(8468) allows the identification of metal-rich giants and their separation from supergiants.

For spectral types later than $\sim$M4, the measurement of this feature becomes very difficult, as it is located inside the TiO $\lambda\lambda$~8432-8442-8452 feature. Indeed, Fig.~\ref{blend} shows high dispersion with luminosity class for stars later than M2, the spectral type at which this TiO feature starts to become strong. As a consequence, the luminosity classification for stars later than M3 can only be achieved by considering the relative strength of the whole metallic spectrum.

\subsection{Radial velocities}

Cross-correlation, either with a model spectrum or a star with known radial velocity $v_{{\rm rad}}$, is the most effective way to derive radial velocities via the Doppler shift introduced in the spectrum by motion in the radial direction. If the template spectrum is chosen to match closely the stellar spectrum, accuracies in $v_{{\rm rad}}$ as high as $\la10$\% of the spectral resolution can be obtained for spectra including a large number of lines.  

We tried two different approaches to derive radial velocities. Initially, we employed direct cross-correlation between our spectra utilising the {\tt hcross} task within 
{\sc 
figaro}. This allows direct comparison between our targets. Radial velocities may be measured relative to members of Ste~2, whose velocities in the Local Standard of Rest (LSR) have been accurately measured by \citetalias{davies07}.

Unfortunately, in the spectral range that we have observed, the spectrum of late giants and supergiants changes drastically with $T_{{\rm eff}}$ due to the appearance of strong TiO bands. As a consequence, we need to use as templates a set of spectra that can match our heterogeneous dataset. The stars in Ste~2 do not cover the whole range of spectral types of our targets.

In view of this, we decided to use again the spectral library of \citet{cenarro01}, taking advantage of the fact that the spectra cover the whole parameter space in 
$T_{{\rm eff}}$ and luminosity at a resolution comparable to ours. In addition, the spectra in the library have been shifted to a common velocity.

The algorithm that obtains the Doppler shift is a two-step process. It starts with a first estimation of $T_{{\rm eff}}$ via the TiO calibration presented in Section~\ref{sec:teff}, which is then used to select a template for comparison from the library. Successive velocity shifts, in steps of 1$\:{\rm km}\,{\rm s}^{-1}$, are applied to the template, and the correlation maximum gives us a first estimation of $v_{{\rm rad}}$. Using this value, we shift the observed spectrum to the rest velocity, and then a second $T_{{\rm eff}}$ is derived. This second estimate is in most cases within $\pm1$ spectral types from the first one. Although a new template could be selected, such a small change in spectral type has a negligible effect on the derivation of $v_{{\rm rad}}$.
Observed $v_{{\rm rad}}$ were transformed into the LSR reference system using
IRAF's\fnmsep\footnote{IRAF is
  distributed by the National Optical Astronomy Observatories, which
  are operated by the Association of Universities for Research in
  Astronomy, Inc., under cooperative agreement with the National
  Science Foundation} {\it rvcorrect} package.

The measured values of $v_{{\rm rad}}$  depend on the spectral range chosen. Because of the presence of strong telluric bands, wide regions of the spectrum cannot be used 
in the cross-correlation. As the humidity  was high during the observations, we restricted the spectral range used for cross-correlation to 8370--8900\AA.  In 
Fig.~\ref{muestra}, we display the sky transmission from \citet{hinkle}. The strongest atmospheric features in this range have 5\% of the intensity of the O$_{2}$ band centred at 8231\AA. By scaling against the 8231\AA\ band, we estimate that the strongest atmospheric features reach a depth of $\la0.03$ continuum units, while the typical metallic lines in the spectrum are close to 0.1 continuum units. Therefore no atmospheric features are noticeable in our spectra and they should not contribute significantly to the cross-correlation function.

Likewise, interstellar features in this range are not numerous. The only strong interstellar feature is the 8620\AA\ diffuse interstellar band (DIB), discussed in the next section. There is also a weaker DIB at 8763\AA\ \citep[e.g.,][]{gredel01} and a broad shallow DIB around 8648\AA\ \citep{munari08,negwd1}. Finally, interstellar lines corresponding to the  (2,0) band of the C$_{2}$ Phillips system are seen in stars with high reddening \citep[e.g.,][]{gredel01}. The strongest features, such as the Q(2) and Q(4) transitions at $\lambda\lambda$8761, 8764\AA, reach intensities of a few hundredths of the continuum in the spectrum of Cyg~OB2 \#12 \citep[$A_{V}\approx10$;][]{gredel01}, but are not seen in any of our spectra. This lack of detectability suggests that the cross-correlation is not affected by the presence of interstellar features.

We derived an initial estimate of the accuracy of the measurements by performing cross-correlations against different templates and among our targets. From this we 
estimated the {\it 
internal} error of the procedure to be around $\pm5\:{\rm km}\,{\rm s}^{-1}$.

\section{Results}
\label{sec:res}

A quick look  at the spectra obtained (Figs.~\ref{fig:rsgc2} to~\ref{fig:al9spect}) clearly shows that all the stars selected by our photometric criteria are luminous 
late-type objects. Accurate classification can be accomplished using the criteria explored in the previous section and direct comparison to the standard stars in the 
library of \citet{cenarro01}. 
To determine luminosity, we use the EW of the \ion{Ca}{ii} triplet lines and the 8468\AA\ blend, as well as the direct comparison of the full metallic spectrum against standard stars in \citet{cenarro01} and \citet{carquillat97}. For stars later than M3, the determination of luminosity is difficult, and relies solely on the apparent strength of the metallic spectrum.

In addition to spectral classification, two other spectral features are important for the analysis. Firstly, the radial velocity of the objects can provide information about their distance. As shown by \citet{davies08}, the Galactic rotation curve allows the separation of stars located in different Galactic arms. In this direction, the line of sight cuts the Sagittarius Arm at $\sim1.5$~kpc from the Sun (where $v_{\rm LSR}$ values are $\la 25\:{\rm km}\,{\rm s}^{-1}$) and crosses the Scutum-Crux Arm at a distance $\sim$3.5~kpc \citep{turner80}, where $v_{\rm LSR}$ values should be $\sim 50\:{\rm km}\,{\rm s}^{-1}$.  Then it reaches the structure close to the edge of the Long
  Bar, perhaps touching it at $\sim$5~kpc and crossing it for 2 or 3~kpc. It does not reach any other arm until it cuts again the  
  Sagittarius Arm at $\ga$10~kpc. Stars in RSGC1 have an average $v_{{\rm
    LSR}}=120\:{\rm km}\,{\rm s}^{-1}$ \citep{davies08}, while those in Ste~2 have $v_{{\rm LSR}}=110\:{\rm km}\,{\rm s}^{-1}$ \citepalias{davies07}. Stars in the Scutum Complex thus have much higher radial velocities than foreground or background objects.

Secondly, the DIB centred at 8620\AA\ has been found to be an excellent tracer of reddening. For moderate reddenings, a very tight correlation $E(B-V) = 2.72(\pm0.03) \times~$EW(DIB)~(\AA) has been found \citep{munari08}. The DIB does not show any correlation with reddening for stars in the heavily reddened ($A_{V}=10.8$) open cluster Westerlund~1 \citep{negwd1}. OB supergiants in this cluster show EW(DIB)$=1.1\pm0.1$\AA, suggesting that the band saturates at this value. Such a strong feature is easily recognisable in the spectra of heavily reddened stars and serves as an additional indicator of high distance. In RSGs, it appears blended with the \ion{Fe}{i}~8621\AA\ line (see Fig.~\ref{muestra}). Because of this, its EW cannot be measured accurately at our resolution. However, the presence of this feature can be deduced by comparison to \ion{Fe}{i}~8611\AA. This line is always stronger than the \ion{Fe}{i}~8621\AA\ feature at all spectral types. Observing a ratio \ion{Fe}{i}~8621\AA/\ion{Fe}{i}~8611\AA$>$1 implies the presence of a moderately strong DIB blended with the \ion{Fe}{i}~8621\AA\ feature.  

\subsection{Stephenson~2}

\begin{figure}
\resizebox{\columnwidth}{!}{
\includegraphics[angle=0,clip]{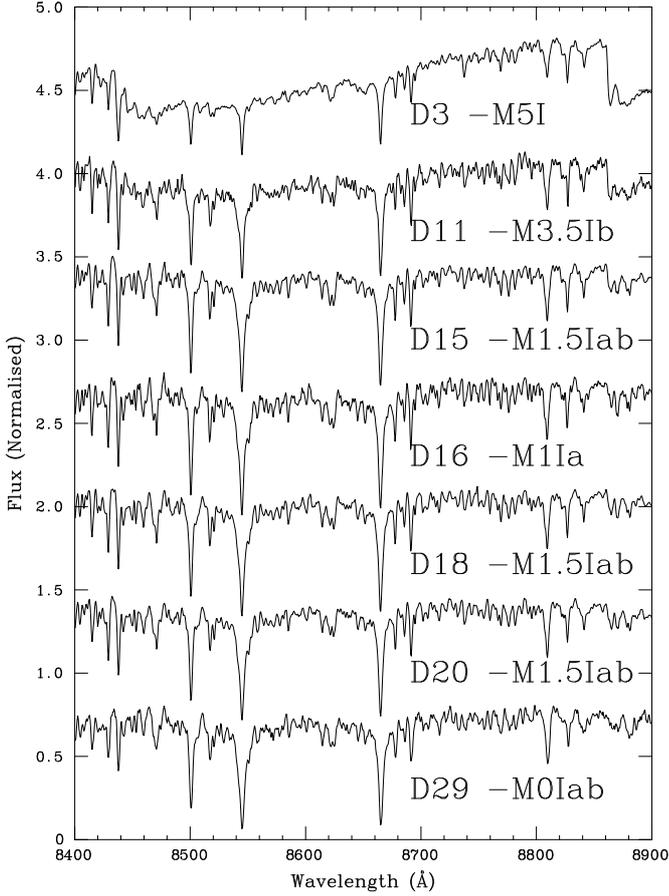}}
\caption{Spectra of seven red supergiants in Ste~2. Note the prominent DIB at 8620\AA\ compared to the neighbouring \ion{Fe}{i}~8611\AA. The \ion{Fe}{i}~8621\AA\ line can be seen as a bump on the red side of this feature, because it is displaced to $\lambda$8624 due to the high $v_{{\rm rad}}$ of these objects, except in D3, which is too late to show this feature.\label{fig:rsgc2}}
\end{figure}

We obtained spectra of seven stars listed as members of Ste~2 by \citetalias{davies07}. Their spectra are displayed in Fig.~\ref{fig:rsgc2}. All the objects are M-type luminous stars. In Table~\ref{tab:ste2}, we list their spectral types as estimated from the strength of the CO 2.3$\mu$m bandhead in \citetalias{davies07}, as estimated using the strength of the TiO 8660\AA\ band and as derived from direct comparison to standard stars.

The agreement in general is very good, except for D18, which was classified M4 by \citetalias{davies07} and appears rather earlier here. The observations of \citetalias{davies07} were taken in 2006, three years before our run. Some RSGs are known to show spectral variability, though this is a rare occurrence, perhaps confined to the most luminous objects \citep{levesque, clark10}.

\begin{table*}
\caption{New spectral types for members of Stephenson~2.\label{tab:ste2}}
\begin{tabular}{lccccccccccccccc}
ID &   \multicolumn{2}{c}{Co-ordinates}  & $i$ &$J$ & $H$    & $K_{{\rm S}}$ & Spectral &  Spectral & Spectral&$v_{{\rm LSR}}$\\
   &   RA     &    Decln.  &   &&&& Type (CO)& Type (TiO) &Type&(km\,s$^{-1}$)\\
\hline
\hline
D3  & 18 39 24.6 &$-$06 02 13.8&12.85&7.27 &5.46 &4.50 & M4\,I&M6 &M5\,Iab &+101\\
D11 & 18 39 18.3 &$-$06 02 14.3&14.35&8.35& 6.21 &5.26 & M4\,I&M3 &M3.5\,Ib  & +106\\
D15 & 18 39 22.4 &$-$06 01 50.1&13.07&8.13& 6.35 &5.51 & M2\,I&M2 &M1.5\,Iab & +99\\
D16 & 18 39 24.0 &$-$06 03 07.3&13.20&8.24& 6.44 &5.60 & M3\,I&M1 &M1\,Ia& +97 \\
D18 & 18 39 22.5 &$-$06 00 08.4&12.88&8.18& 6.45 &5.63 & M4\,I&M1 &M1.5\,Iab & +101\\
D20 & 18 39 24.1 &$-$06 00 22.8&13.18&8.43& 6.66 &5.81 & M2\,I&M1 &M1.5\,Iab& +103\\
D29 & 18 39 22.2 &$-$06 02 14.7&13.17\tablefootmark{a}&8.61& 6.88 &6.15 & M0\,I&M1 &M0\,Iab & +107 \\
\hline
\end{tabular}
\newline\\
\tablefoottext{a}{The DENIS magnitude seems to correspond to a blend of two stars, D29 and 2MASS J18392193$-$0602165.}
\end{table*}

\citetalias{davies07} find an average radial velocity for the cluster $v_{{\rm LSR}}=109.3\pm0.7\:{\rm km}\,{\rm s}^{-1}$, with the uncertainty representing Poisson statistics for 26 members with measured values. The seven stars that we have observed have average $v_{{\rm LSR}}=109\pm2\:{\rm km}\,{\rm s}^{-1}$ (where the error is the standard deviation of the seven measurements), i.e., are representative of the cluster velocity. Their individual values range from $105\:{\rm km}\,{\rm s}^{-1}$ for D29 to $112\:{\rm km}\,{\rm s}^{-1}$ for D15.

When we compare our values, we find that the differences in $v_{{\rm rad}}$ (D07 minus our values) range from $+13\:{\rm km}\,{\rm s}^{-1}$ to $-2\:{\rm km}\,{\rm s}^{-1}$, with an average difference of $+8\:{\rm km}\,{\rm s}^{-1}$ and a standard deviation of $5\:{\rm km}\,{\rm s}^{-1}$. For D3, we have three measurements, which give values of 100, 111 \& $93\:{\rm km}\,{\rm s}^{-1}$, corresponding to an average of $101\pm9\:{\rm km}\,{\rm s}^{-1}$.

Based on these values, we estimate that the total error in one single measurement of $v_{{\rm rad}}$ using our technique is  $\pm10\:{\rm km}\,{\rm s}^{-1}$ or slightly higher\fnmsep\footnote{This value is slightly higher than expected. However, we note that it is not due to use of external templates, as direct crosses between the three spectra of D3 using {\tt hcross} give similar dispersion. We can only speculate about the reason why the error is larger than the expected 10--15\% of the spectral resolution. It may be related to the fact that we tried to align several stars in each slit, meaning that objects are occasionally not well centred, perhaps resulting in different light paths for starlight and the arc lamp light.}. However, the technique appears robust when applied to a sample of objects in a cluster. We assume that the difference of  $+8\:{\rm km}\,{\rm s}^{-1}$ represents our systematic shift with respect to the LSR system adopted by \citetalias{davies07}, based on the radial velocity of Arcturus. The average radial velocity that we obtain for Ste~2 is  $v_{{\rm LSR}}=102\pm4\:{\rm km}\,{\rm s}^{-1}$. The difference with the average value of \citetalias{davies07} is  $+7\:{\rm km}\,{\rm s}^{-1}$, consistent with a systematic shift.

 The reason for this shift is unclear. It may be due to a different definition of the LSR. {\it rvcorrect} transforms using the Standard Solar Motion ($+20\:{\rm km}\,{\rm s}^{-1}$ towards $\ell=56\degr$, $b=23\degr$). We have calculated mean velocity corrections for the stars observed in Ste~2 $\Delta V_{\sun}=-4.5\:{\rm km}\,{\rm s}^{-1}$ and $\Delta V_{{\rm LSR}}=+16.0\:{\rm km}\,{\rm s}^{-1}$. In any case, the shift is of the same order of magnitude as the dispersion in our measurements for cluster members, and so could be statistical, as the number of members observed is low. In addition to the accuracy of individual measurements, a real dispersion in $v_{{\rm rad}}$ amongst cluster members could be present. Observations of five RSGs in the association surrounding $\chi$~Persei show random fluctuations in radial velocity with amplitudes $\ga 5\:{\rm km}\,{\rm s}^{-1}$ \citep{mermilliod}.

\subsection{RSGC3}

\begin{figure}
\resizebox{\columnwidth}{!}{
\includegraphics[angle=0,clip]{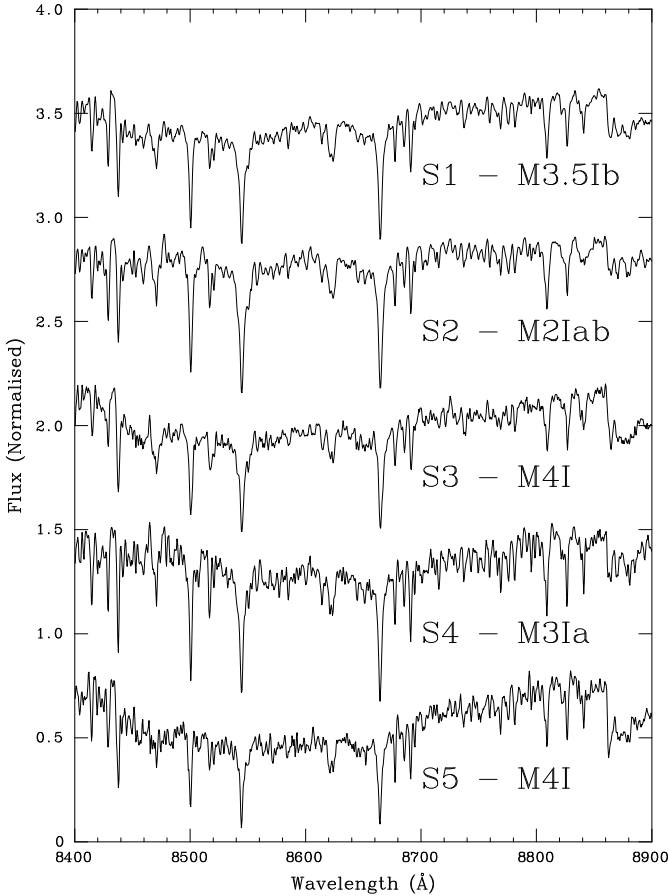}}
\caption{Spectra of five red supergiants in RSGC3. Note the prominent DIB at 8620\AA\ compared to the neighbouring \ion{Fe}{i}~8611\AA.\label{fig:cluster3}}
\end{figure}

We observed five stars in RSGC3, those numbered S1--5 in \citet{clark09}. Their spectra are displayed in Fig.~\ref{fig:cluster3}. The five objects are M-type supergiants. Their measured $v_{{\rm rad}}$ are consistent within the accuracy discussed above (S5 shows the value deviating most clearly, but it also has the poorest spectrum with SNR$\sim$35 per resolution element). The average velocity for these five members is $v_{{\rm LSR}}=+95\pm7\:{\rm km}\,{\rm s}^{-1}$, where the error represents the standard deviation.  Interestingly, \citet{alexander09} find that the CO distribution in this direction shows a maximum at $v_{{\rm LSR}}=+95\:{\rm km}\,{\rm s}^{-1}$. Material with this $v_{{\rm LSR}}$ shows a spatial distribution strongly correlated with the GLIMPSE
$8\:\mu$m emission, and presents a hole around the cluster, suggesting a cavity blown by stellar winds and supernova explosions. The  $v_{{\rm LSR}}$ for RSGC3 is smaller than that found for Ste~2 ($v_{{\rm LSR}}=102\pm4\:{\rm km}\,{\rm s}^{-1}$), though consistent within the errors.

\begin{table*}
\caption{Properties of confirmed members of RSGC3.\label{tab:rsgc3}}
\begin{tabular}{lccccccccccccccc}
ID & \multicolumn{2}{c}{Co-ordinates}  &  $I$\tablefootmark{a} &$J$\tablefootmark{a} & $H$\tablefootmark{a}    & $K_{{\rm S}}$\tablefootmark{a} & Spectral & Spectral &Spectral& $E(J-K_{{\rm S}})$ &$v_{{\rm LSR}}$\\
   &      RA     &    Decln.  &&&&& Type (CO)&Type (TiO) &Type&&(km\,s$^{-1}$)\\
\hline
\hline
S1 & 18 45 23.60 &$-$03 24 13.9 & 13.5 & 8.55& 6.54 &5.58  &   RSG & M3 & M3.5\,Ib & 1.80  &+97 \\
S2 & 18 45 26.54 &$-$03 23 35.3 & \tablefootmark{b}&8.53& 6.62 &5.75  &    M3\,I & M2 &M2\,Ia& 1.68  &+104\\
S3 & 18 45 24.34 &$-$03 22 42.1 & 13.7\tablefootmark{c} &8.54& 6.43 &5.35  &      M4\,I & M4&  M4\,I &2.02  & +99\\
S4 & 18 45 25.31 &$-$03 23 01.1 & 13.7\tablefootmark{c}& 8.42& 6.39 &5.31  &      M3\,I & M2& M3\,Ia &1.97  &+93\\
S5 & 18 45 23.26 &$-$03 23 44.1 & 13.8& 8.51& 6.52 &5.52  &      M2\,I & M4& M4\,I&1.82  & +84\\
\hline
\end{tabular}
\newline\\
\tablefoottext{a}{$JHK_{{\rm S}}$ magnitudes from 2MASS. $I$ magnitudes from USNO-B1.0.}
\newline
\tablefoottext{b}{No USNO-B1.0 source, but the star is similarly bright to the other four members observed in the DSS2 $I$ image.}
\newline
\tablefoottext{c}{The magnitude corresponds to a blend of two stars in the DSS2 $I$ image.}
\end{table*}

With these new spectral types, we calculate individual colour excesses $E(J-K_{{\rm S}})$, using the calibrations of \citet{levesque} to calculate the intrinsic  $(J-K)_{0}$ for each spectral type\fnmsep\footnote{Note that the $(J-K)_{0}$ colours obtained by applying the calibration of $T_{{\rm eff}}$ versus spectral type in \citet{levesque} to their $(J-K)_{0}$/$T_{{\rm eff}}$ relationship are almost identical to the empirical calibration of intrinsic $(J-K)_{0}$ for red giants of \citet{straizys09}. This agreement not only gives support to both calibrations, but indicates that the $(J-K)_{0}$ colours of red giants and supergiants are indistinguishable.}. The values derived, shown in Table~\ref{tab:rsgc3}, are very similar to those of \citet{clark09}. Given the relatively good agreement between our spectral types and those derived using the CO-bandhead calibration, changes in the individual absolute magnitudes are small. Errors in $M_{K}$ are dominated by the uncertainty in the distance to the cluster.

\subsection{Field 1}

In Field 1, we selected 12 candidates (listed in Table~\ref{tab:fields}) and  managed to observe seven of them. All the stars observed turn out to be M-type luminous stars. 
Two 
objects, F1S01 and F1S10, have relatively low $E(J-K_{{\rm S}})$ and show no evidence for heavy extinction in their spectra, as the 8620\AA\ DIB is not prominent. Both are 
quite late and their exact luminosity classes are difficult to assign, but both look rather luminous, close to being low-luminosity supergiants.

\begin{figure}
\resizebox{\columnwidth}{!}{
\includegraphics[angle=0,clip]{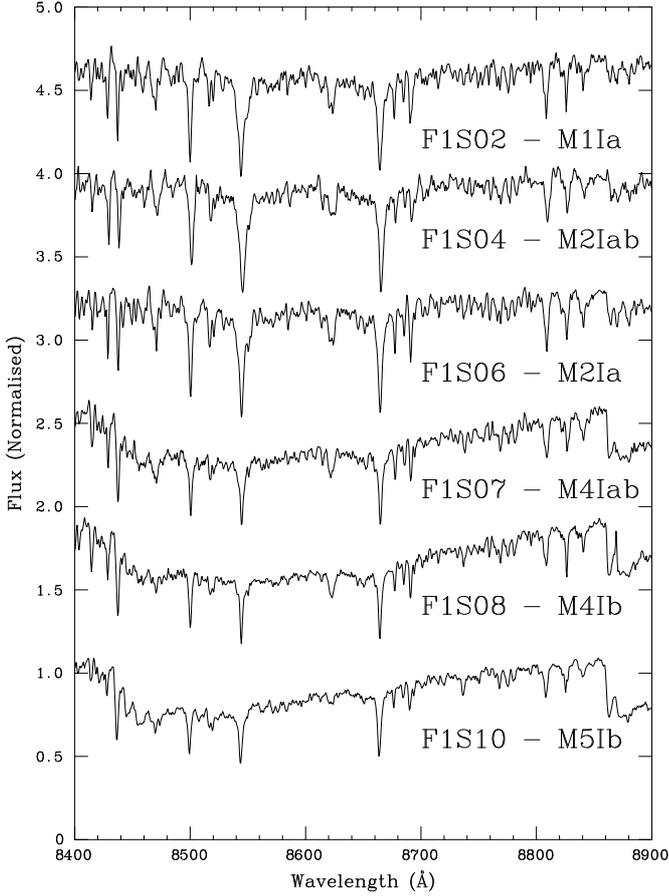}}
\caption{Spectra of red supergiants in Field~1. The five top objects have similar colours and are affected by heavy reddening. The bottom spectrum corresponds to S10, which is less luminous than the other objects. It has much lower reddening than the other stars (as shown by the weak 8620\AA\ DIB) and very different radial velocity.\label{fig:al7spect}}
\end{figure}

The other five objects have similar values of $E(J-K_{{\rm S}})$, all within the same range as stars in RSGC3. As seen in Fig.~\ref{fig:al7spect}, all five show very prominent 8620\AA\ DIB, with EW$\approx$1.3\AA. The feature is so broad that it is almost blended with \ion{Fe}{i}~8611\AA. All five objects have high observed $v_{{\rm rad}}$, but our measurements show higher dispersion than those for stars in Ste~2 and RSGC3.  The average of their velocities is  $v_{{\rm LSR}}=+89\pm16\:{\rm km}\,{\rm s}^{-1}$, where the error represents again the standard deviation. All the individual measurements are within one standard deviation, but deviate slightly more than the estimated error of $\pm10-12\:{\rm km}\,{\rm s}^{-1}$ (all the spectra are moderately poor with typical SNR$\sim$50).

Among the objects that we did not observe, F1S03 with $(J-K_{{\rm S}})=1.89$ and F1S11 with $(J-K_{{\rm S}})=1.79$ must have relatively low colour excesses if they are K--M stars. On the other hand, F1S05 and F1S09 have colours similar to the group of five supergiants. Below we argue that all these high-reddening objects represent a new cluster of red supergiants.
 
\subsection{Field 2}

There is an obvious concentration of bright stars in the $K_{{\rm S}}$ image of the area. Only 11 of them pass our photometric cut, defining an apparent compact group around (RA: 18:44:45, $\delta$: $-$03:14:30) and a smaller group $\sim6\arcmin$ to the North. We find seven candidates in the main group and four in the northern group. Their coordinates and magnitudes are listed in Table~\ref{tab:fields}.

We observed five candidates in the main group. Of these, three objects F2S01, F2S03 and F2S04 are M-type supergiants with similar colours. Their colour excesses are almost 
identical, and similar to the lowest values displayed by stars in RSGC3. Their observed $v_{{\rm rad}}$ are also similarly high. However, there is no strong evidence for a large cluster here. F2S05 and F2S07 are both late-M giants. Their colour excesses are much lower, even if their $v_{{\rm rad}}$ are high. Of the stars that we did not observe, F2S06 has $(J-K_{{\rm S}})=1.56$, and cannot be highly reddened if it is a red star. F2S03, on the other hand, has 
$(J-K_{{\rm S}})=3.46$, and it could be a redder or more highly reddened supergiant. 

\begin{figure}
\resizebox{\columnwidth}{!}{
\includegraphics[angle=0,clip]{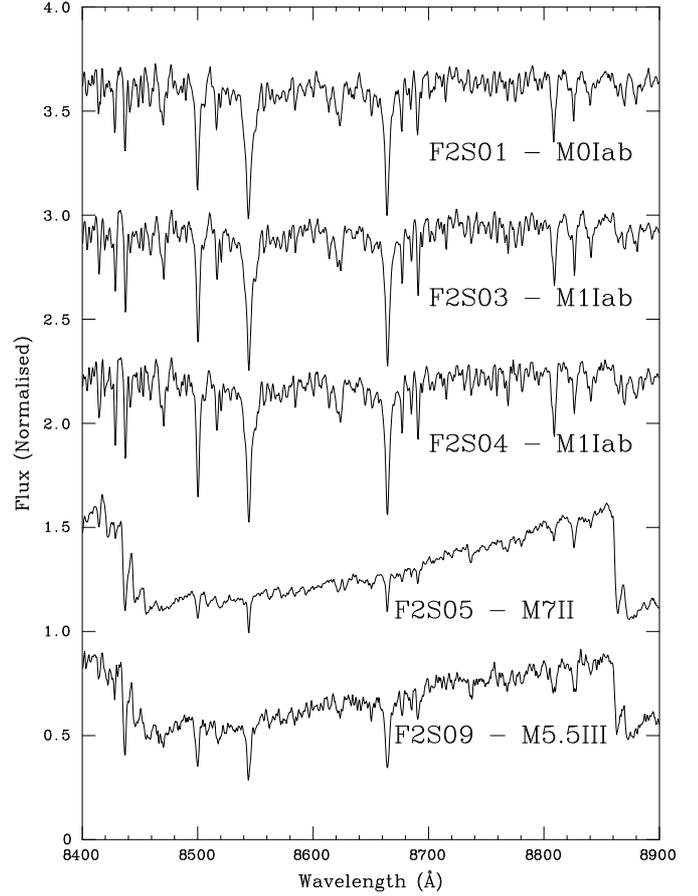}}
\caption{Spectra of stars in Field~2. The three top objects have similar colours and are affected by heavy reddening. They are likely members of the RSGC3 association. F2S05 is a luminous foreground giant. F2S09  morphologically looks like a giant (though the spectrum has low SNR), but its reddening and $v_{{\rm rad}}$ are similar to those of supergiants in the RSGC3 association.\label{fig:al9spect}}
\end{figure}

In the northern group, F2S09 has a late spectral type. The spectrum is quite noisy, and the luminosity class is difficult to determine. It looks like a giant (see Fig.~\ref{fig:al9spect}), but displays high reddening and high $v_{{\rm rad}}$, similarly to the supergiants. F2S11, which we did not observe, has colours and magnitudes similar to the three supergiants in the main group.

\subsection{Field 3}

We observed six of the ten candidates selected in this area (Table~\ref{tab:fields}). All of them are mid or late M giants, with moderate reddening. Though their colour excesses $E(J-K_{{\rm S}})$ are quite uniform, they display a very wide range of observed $v_{{\rm rad}}$.

The four stars that we did not observe have redder colours than these giants, with F3S04 and F3S07 showing 2MASS colours similar to members of RSGC3.

\section{Discussion}
\label{sec:discu}

We have observed a sample of RSG members of the massive clusters Ste~2 and RSGC3 in the region of the \ion{Ca}{ii} triplet. In addition, we have observed 19 objects within $20\arcmin$ of RSGC3, selected because their 2MASS colours suggested that they might be reddened luminous late-type stars.
 
\subsection{Validity of criteria and calibrations}

 We have selected targets based on their infrared  brightness and colours. The criterion $0.1\le Q\le0.4$ was expected to reject early-type stars and red dwarfs. 
Indeed, all the 19 objects selected according to  these criteria for which we have spectra turn out to be luminous M-type stars with different amounts of reddening. 
Eight objects are highly reddened supergiants. The rest  are mid and late M giants or low-luminosity supergiants with different amounts of reddening, ranging from 
$E(J-K_{{\rm S}})=0.55$ for F1S01 to 1.36 for F2S09.  Therefore we may conclude that the criteria defined in Section~\ref{sec:2mass} to identify reddened luminous 
late-type stars are very successful, as all the objects selected are M supergiants or late M giants. The selection in $Q$ effectively removes most (perhaps all) low-mass red stars, which are expected to have $Q>0.4$. The spectral characteristics of the objects selected suggest that their lower $Q$ values  (lying in the region corresponding to yellow stars) are likely due to the existence of extended atmospheres, which modify their spectral energy distribution. As the criteria used are mainly related to intrinsic stellar properties rather than to the characteristics of the area surveyed, these criteria may be effective at selecting obscured red luminous stars in other regions.

Unfortunately, the photometric properties of luminous red giants and RSGs seem to be identical, and a separation of both classes is not possible based solely on 2MASS photometry. With the small sample observed here, it is not possible to assess if a combination of 2MASS and {\it Spitzer} photometry can effectively separate the two classes. Though our sample shows some evidence that this may be the case, objects enshrouded in dusty envelopes will necessarily display very red colours caused by dust emission, independently of their luminosity. A larger sample of objects with accurate spectral types will thus be needed to assess this issue.

 A second important result concerns the use of the CO bandhead features in the $K$ band to estimate the spectral type. Different authors have calibrated the EW of the bandhead against spectral type, finding a degree of confusion between giants and supergiants, and an estimated uncertainty of $\pm 2$ spectral subtypes for a given star \citep{davies07,alexander09,neg10}. Our targets include 11 red supergiants in Ste~2 and RSGC3 for which previous authors had estimated the spectral type using the calibration of the CO bandhead. If we compare the spectral types estimated with those derived by direct comparison to the standards, the agreement is excellent. The average difference is one spectral type. Since the classification of the standard stars was originally carried out in the blue-violet region (meaning that the assignment of spectral types based on $Z$ band spectra is indirect), the agreement is surprisingly good. Even though a given star may have its spectral type misestimated by two subtypes, there does not seem to be any systematic effects.

Finally, for the 31 stars in our sample, we can also compare the spectral types derived from the calibration of the TiO band described in Section~\ref{sec:teff} with those derived by direct comparison to the standards. In this case, the average difference is only 0.6 spectral types, confirming that this calibration provides valid spectral typing for M-type giants and supergiants.

\subsection{A new cluster: Alicante~7 = RSGC5}
\label{sec:al7}

\begin{figure}
\resizebox{\columnwidth}{!}{
\includegraphics[angle=0,clip]{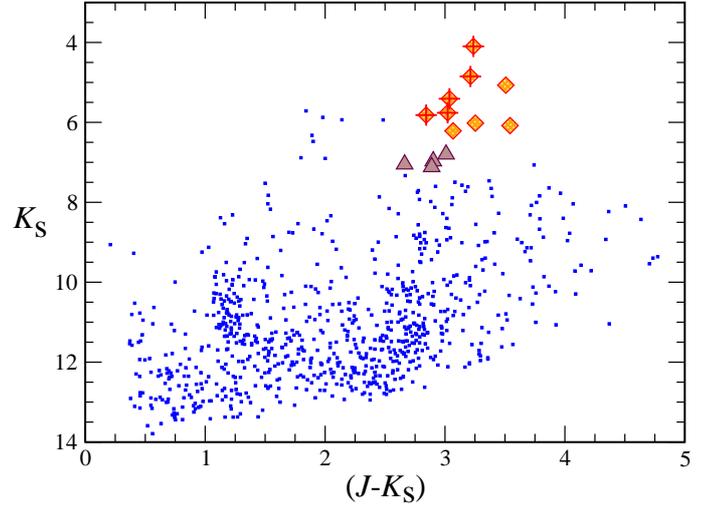}}
\caption{2MASS colour magnitude ($J-K_{{\rm S}}$)/$K_{{\rm S}}$ plot for stars within $7\arcmin$ of the centre of the new cluster Alicante~7. The diamonds identify stars with colours and $Q$ in the same range as RSGs in RSGC3. Five of them, marked by crosses, are spectroscopically confirmed as M-type supergiants. Triangles are other stars in the area with very similar colours, but fainter magnitudes. Note that the brightest star, A6 in Table~\ref{tab:props}, is saturated in $K_{{\rm S}}$. \label{fig:al7hr}}
\end{figure}

Since five RSGs detected in Field~1 seem to have very similar colours, we investigated the possibility that they comprise a new cluster of RSGs. To accomplish  this, we 
undertook a similar
photometric analysis similar to that presented in  \citep{clark09} for RSGC3. We took a $7\arcmin$ circle around a nominal centre, chosen as the position of F1S06 and 
selected 2MASS photometry 
for stars with quality flags "AAA" and $\Delta K_{{\rm S}}\leq0.05$\fnmsep\footnote{Note that S8 does not fulfil these conditions, as it is saturated in $K_{{\rm S}}$. Nevertheless, we include it in the sample, as it is certainly a red supergiant.}.

\begin{figure}
\resizebox{\columnwidth}{!}{
\includegraphics[angle=-0, clip]{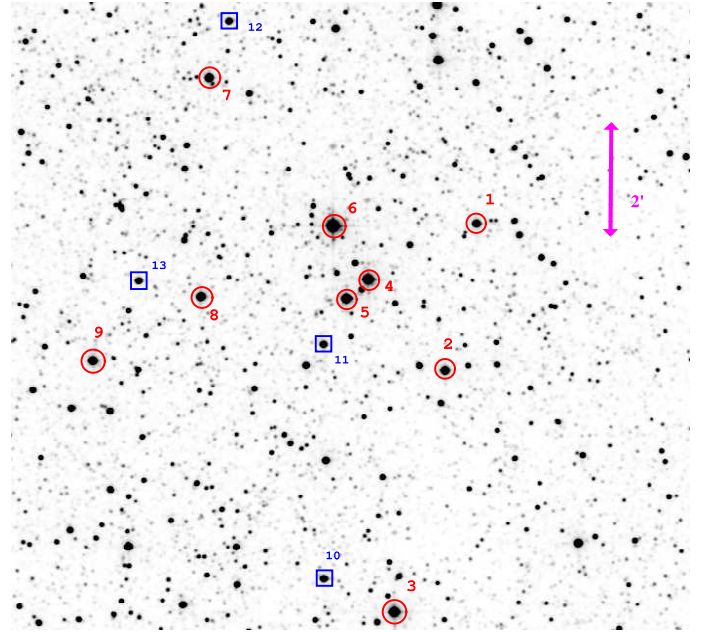}}
\caption{Finding chart for Alicante~7, with the stars listed in Table~\ref{tab:props} indicated. The finder comprises a $K$-band image from 2MASS with a $\sim11\arcmin\times11\arcmin$ field of view centred on the position of A5 (=F1S06 in Table~\ref{tab:fields}). Red circles indicate likely RSG members (including the five objects spectroscopically confirmed), while blue squares mark stars with similar colours, but fainter magnitudes.\label{al7finder}}
\end{figure}

We found  a number of bright objects, clearly grouped in the ($J-K_{{\rm S}}$)/$K_{{\rm S}}$ and  ($H-K_{{\rm S}}$)/$K_{{\rm S}}$ diagrams, which stand up from the rest of 
the field  (see Fig.~\ref{fig:al7hr}). This group includes all the five M-type supergiants identified, plus F1S09 and three other stars with similar colours and the expected $Q$, but faint 
$i$ magnitudes. In addition, we found four other stars with similar colours but fainter magnitudes. This second group is not so clearly separated from the field population 
as the nine main candidates.  We identify these nine main candidates as a new cluster of red supergiants, which we provisionally name Alicante~7. Figure~\ref{al7finder} shows a finder for all the candidates.

Star A6 (=F1S08) is the brightest cluster member in the $K_{{\rm S}}$ band. Its $K_{{\rm S}}$ magnitude is in the saturation range for 2MASS. This object is a strong mid-IR source,  and almost certainly the counterpart to IRAS~18418$-$0331. This suggests that it may be surrounded by a dusty envelope. If this is the case, its luminosity class, based on the strength of the metallic spectrum, may be incorrect, as the spectrum may be veiled by dust, leading to the dilution of the metallic features.

\begin{figure*}[ht!]
\resizebox{\textwidth}{!}{\includegraphics[angle=90,clip]{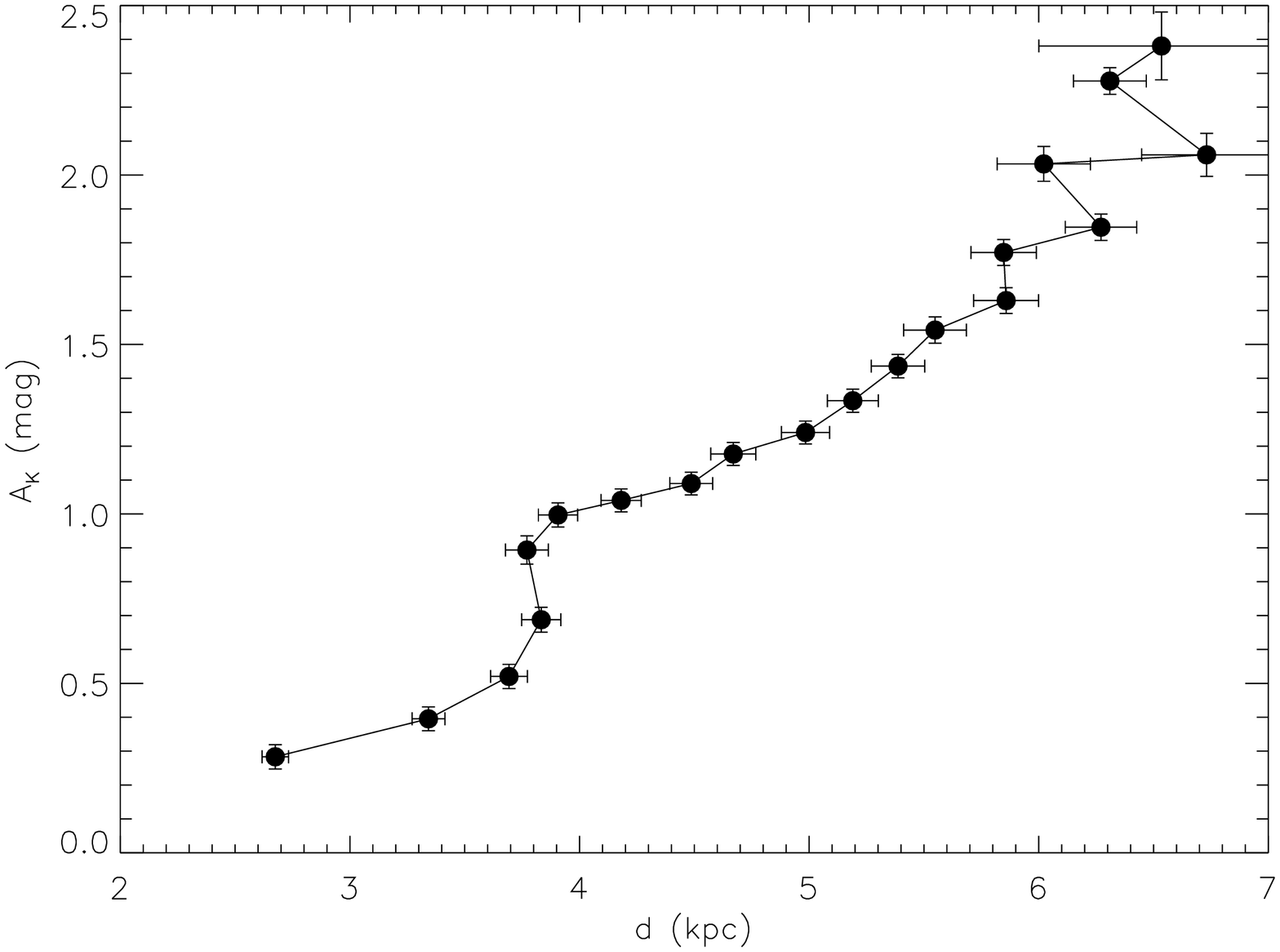}
\includegraphics[angle=90,clip]{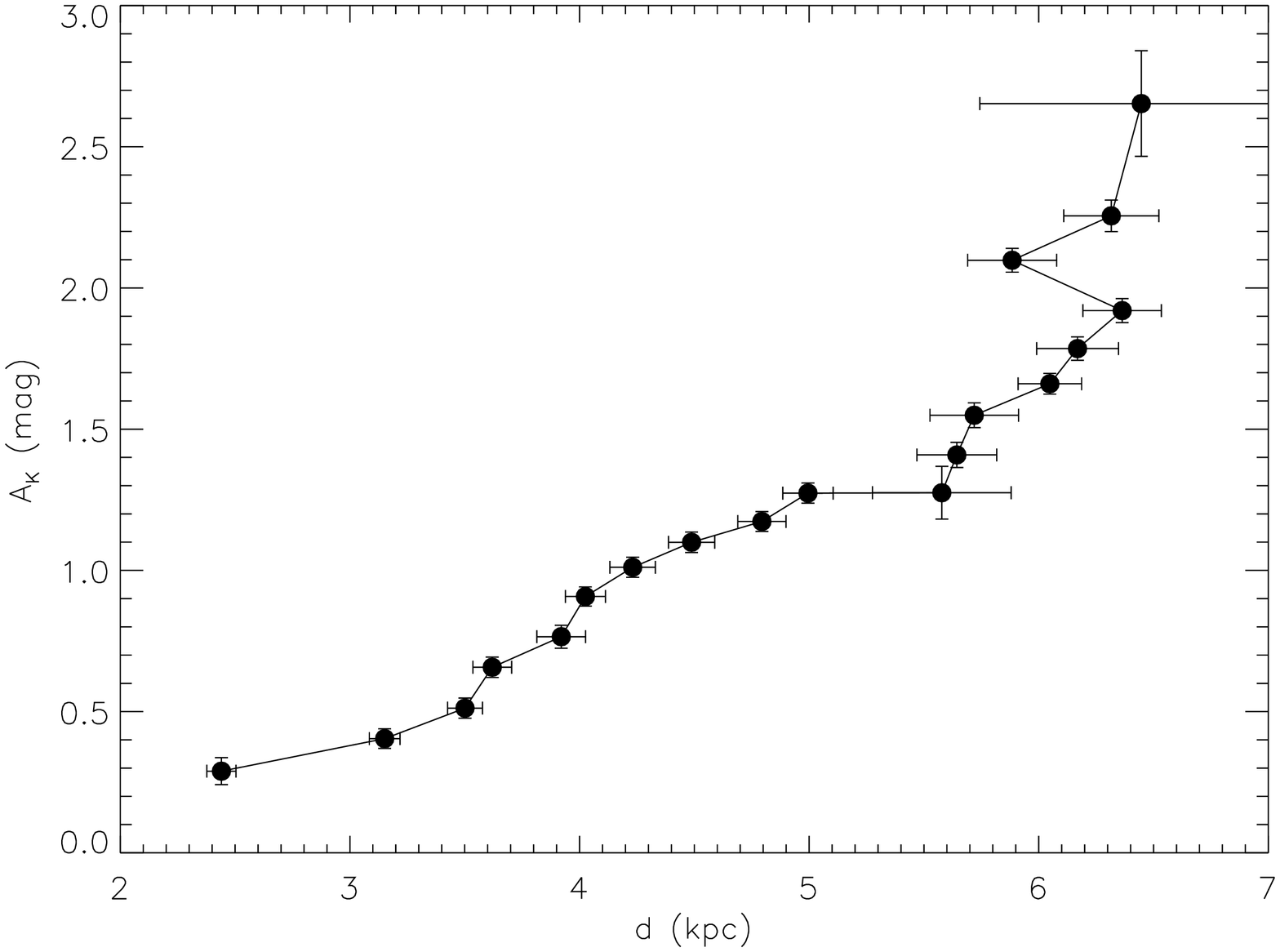}}
\caption{{\bf Left panel: }Run of the extinction in the direction to Alicante~7. The data have been obtained by applying the technique of \citet{cabrera05} to the red clump giant population within $20\arcmin$ from the nominal cluster centre. {\bf Right panel: } Same for the field surrounding RSGC3. Both runs are pretty similar. The sudden increase in the reddening at $\sim 6$~kpc provides a direct estimate on the distance to both clusters.\label{fig:extin}}
\end{figure*}

We can obtain an estimate of the distance  to Alicante~7 by studying the distribution of interstellar extinction in its direction. Following the technique of 
\citet{cabrera05}, we used the  population of red clump giants in infrared colour/magnitude diagrams. As the 2MASS data are not deep enough for this purpose, we used 
UKIDSS data. We plotted the ($J-K$)/$K$ diagram within $20\arcmin$ of the cluster centre, and selected the giant population, which is seen in such plots as a well defined 
strip. We chose a $20\arcmin$ radius  as a compromise that allowed us to construct a representative  $(d,A_K)$ curve  for the cluster sightline, while providing a 
large enough number of 
stars to permit an accurate determination. Decreasing this value to, for example, $10\arcmin$ produces noisier results, but does not change the overall behaviour of the 
extinction. We 
perform the same calculation for the area centred on RSGC3. Given the distance between the two clusters, the two areas partly overlap. The results are shown in Fig.~\ref{fig:extin}. 

As expected, both extinction runs are similar. The extinction starts at low values and suffers an important increase at a distance of 3.5--4~kpc, which is believed to correspond to the position of the first crossing of the Scutum-Crux Arm along this line of sight. Around $\ell=28\degr$, \citet{turner80} found several luminous OB supergiants with distances in excess of 3~kpc and reddenings $E(B-V)\approx1.3$, corresponding to $A_{K}=0.4$, in very good agreement with our determination. After this, the reddening increases steadily, reaching a value $A_{K}\approx1.5$ at 6~kpc. The extinction increases abruptly between 6 and 7~kpc, reaching values that render the technique unusable for higher distances. From this, it is clear that both Alicante~7 and RSGC3, with $A_K\approx1.5$ are located at a distance $\sim6$~kpc, just in front of the extinction wall.  This distance determination is in very good agreement with the dynamical distance found by \citet{alexander09} for the CO material apparently surrounding RSGC3, namely $d=6.1$~kpc.

\subsection{An extended association surrounding RSGC3}
\label{extended}

\begin{figure}
\resizebox{\columnwidth}{!}{
\includegraphics[angle=0,clip]{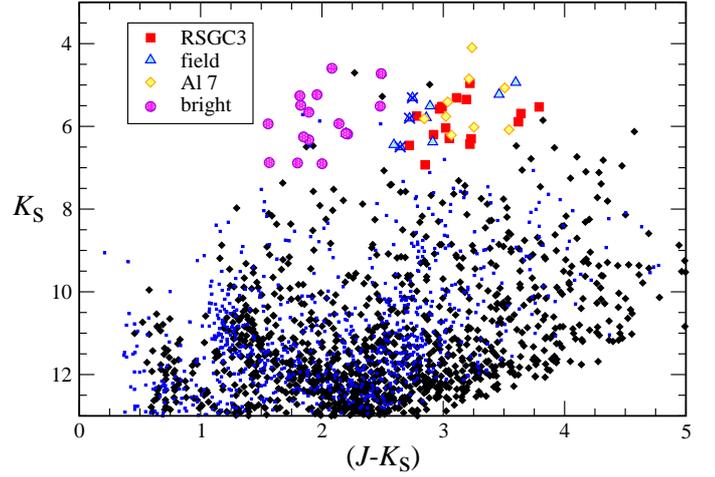}}
\caption{Composite 2MASS colour magnitude ($J-K_{{\rm S}}$)/$K_{{\rm S}}$ plot for the whole area surrounding RSGC3. Black dots are stars within $7\arcmin$ of the centre of RSGC3, while blue microdots are stars within $7\arcmin$ of the centre of Alicante~7. Red squares represent the 16 most likely RSG members of RSGC3, most of which are confirmed spectroscopically. Yellow diamonds as likely RSG members of Alicante~7, as in Fig.~\ref{fig:al7hr}. Blue triangles are all the stars selected in Field~2 and Field~3 with $i>13.2$. The three objects marked with crosses are the three M-type supergiants with high reddening spectroscopically identified in Field~2. Finally, the circles represent all the objects selected in the three fields with $i<13.2$, none of which is a reddened supergiant.\label{fig:combine}}
\end{figure}

 Comparison of Tables~\ref{tab:rsgc3} and~\ref{tab:props} reveals that the RSGs in RSGC3 and Alicante~7 have very similar colours and magnitudes, suggesting that they share a common distance and age. In Fig.~\ref{fig:combine}, we plot the 2MASS ($J-K_{{\rm S}}$)/$K_{{\rm S}}$ diagram for all the stars with high-quality photometry within $7\arcmin$ of the centres of both clusters, together with all  our photometric candidates  from Table~\ref{tab:fields}. Several populations can be seen in both diagrams, revealing the similarity of both sightlines. At 6~kpc, the separation between the two clusters is 24~pc. Therefore the two groups of RSGs must be two separate clusters. Both clusters, however, appear behind approximately the same amount of obscuration, and their RSGs display very similar 2MASS colours and magnitudes. Given that the two clusters are situated at the same distance, the RSGs must have the same intrinsic luminosity, indicating that the two clusters have the same age. 

The age of RSGC3 has been estimated at 16--20~Myr \citep{clark09}. Simulations of stellar populations using a Salpeter IMF suggest that a population of $10\,000\:M_{\sun}$ at this age should contain eight RSGs \citep{clark09}. Other simulations, using a Kroupa IMF, reduce the number to 2--5 RSGs for each $10\,000\:M_{\sun}$ \citep{simonw51}. Based on this, Alicante~7 must contain a minimum of $10\,000\:M_{\sun}$ to have a population of nine RSGs and could be up to $\sim3$ times more massive.

The small group of RSGs found in Field~2 may be a smaller cluster, with four or five supergiants, but this is difficult to demonstrate; we  refer to this {\em putative} cluster 
as 
Alicante~9. As long as the unevolved populations of the clusters in the Scutum 
Complex cannot be detected,  it will be 
impossible to distinguish similarly small clusters  from random concentrations of bright red stars. As a consequence, estimating the total mass of the 
association 
surrounding RSGC3 is very difficult. For example, a cluster with only four RSGs,  which will be very difficult to identify as such, may easily contain $\ga5\,000\:M_{\sun}$. 

In order to make a rough estimate of the population, we consider the population of bright red stars that are not members of the association.
Of the 19 objects in the area surrounding RSGC3 for which we have spectra, 11 are unlikely to be members, as they present much lower extinction than cluster members. Most of them have $E(J-K_{{\rm S}})\approx0.6-0.8$ and relatively low $v_{{\rm LSR}}$. They are likely located at distances $d\sim3.5$~kpc, where the line of sight cuts the Scutum-Crux arm for the first time\fnmsep\footnote{At this distance, the Galactic rotation curve predicts velocities $v_{{\rm LSR}}\approx40-60\:{\rm km}\,{\rm s}^{-1}$ \citep{davies08}, meaning that there is good agreement between reddening and $v_{{\rm LSR}}$ for a number of objects such as F1S01, F1S10, or F3S08. Other objects, such as F3S01 or F3S05, have very discrepant $v_{{\rm LSR}}$}. All these objects are relatively bright in $i$, with typical values in the 11--12 range, in good agreement with the moderately low extinction\fnmsep\footnote{One exception is F2S09, which has high $E(J-K_{{\rm S}})$ and $v_{{\rm LSR}}= 94\:{\rm km}\,{\rm s}^{-1}$. This object is also the faintest amongst the giants in $I$.These values suggest that it must be just in front of RSGC3.}. In contrast, members of RSGC3 and other M-type supergiants in Alicante~7 and Alicante~9 have rather fainter $i$ magnitudes, in most cases $\ga13.5$. 

In view of this, it seems possible to  separate members  of the RSGC3 association from interlopers using their $i$ magnitudes. Based on the observed 
magnitudes of confirmed members and confirmed foreground stars, we may set the dividing line at $i=13.2$. In Fig.~\ref{fig:combine}, we plot all the photometric candidates from the three fields, using different symbols for stars brighter or fainter than $i>13.2$. We find that stars with $i>13.2$ fall inside the area of the diagram occupied by cluster members and confirmed supergiants. In contrast, stars with $i<13.2$ occupy a different region of the diagram, with lower $(J-K)$ colours. The only exception is F2S09, whose nature is unclear. It has high $E(J-K_{{\rm S}})$ and $v_{{\rm LSR}}= 94\:{\rm km}\,{\rm s}^{-1}$. Therefore, if it is a giant, it must be located just in front of the RSGC3 association.

 Since brightness in $i$ and $(J-K_{{\rm S}})$ seem to be well correlated, it seems safe to assume that any star in the vicinity of RSGC3 with $0.1\leq Q \leq 0.4$ and $2.5<(J-K_{{\rm S}})<3.7$ has a very high probability to be a RSG in the RSGC3 association.  If we turn now to the candidate members of RSGC3 listed by \citet{clark09}, we find that 18 of them fulfil all criteria, defining a compact core $\sim3\arcmin$ across and a cluster radius  $\sim6\arcmin$.

 At larger distances, apart from the nine candidate members of Alicante~7, there are $\ga25$ objects fulfilling these criteria within  $r=18\arcmin$ (3 cluster radii) of the cluster core . None of them can be identified with a DENIS or USNO-B1.0 sources at $i<13.2$, but 18 have counterparts fainter than this limit (red circles in Fig.~\ref{finder}). Most of these objects are likely RSGs in the association. If we take into account that criteria are not designed to select {\it all} RSGs and that some objects pass all the 2MASS criteria but have no optical counterpart, the association including RSGC3 must contain $\ga50$ RSGs within $18\arcmin$ ($\sim31$~pc at 6~kpc). Using the estimates of \citet{simonw51}, such a population of RSGs would require an underlying population of at least $10^{5}\:M_{\sun}$. Of course, such a huge structure needs to be confirmed via 
high resolution spectroscopy, which can provide kinematic memberships. 

An interesting parallel is provided by the G305 star-forming region. Two massive clusters (Danks 1 \& 2), separated by only 3~pc, have very similar, but measurably different ages, and are surrounded by a halo of active star formation extending up to distances $\approx$30~pc away \citep{cp04,clark11}. Such a region may evolve over the next $\sim$15~Myr to resemble the association around RSGC3 (though with the mass scaled down by perhaps a factor 4--5). On an even smaller scale, the Cas~OB8 association, in the Perseus Arm, provides a better match for the age of the RSGC3 association. The largest cluster in the area, NGC 663, is relatively massive, containing (at an age 20--25~Myr) five blue supergiants in the core region and at least two RSGs in the halo \citep{marco, pandey,mermilliod}. The smaller cluster NGC~654, which contains one blue and one yellow supergiant, is located $\sim30$~pc away (for an assumed $d=2.4$~kpc; \citealt{marco,pandey}). An even smaller cluster, NGC~659, which does not contain supergiants, is $\sim$23~pc away from NGC 663, while NGC~581, which contains one blue and one red supergiant, is $\sim 65$~pc away. All the clusters in the area have approximately the same age, and there seems to be a number of supergiants at a comparable distance spread between the clusters \citep{humphreys}.

\section{Conclusions}

We have conducted a pilot study of the neighbourhood of RSGC3. We have  proposed a number of criteria that allow the separation of candidate luminous red stars from 2MASS 
photometry and observed a sample of the candidates in the 8300--9000\AA\ range at intermediate resolution. All the objects spectroscopically observed were found to be 
M-type giants or supergiants, confirming the validity of the photometric criteria.
Our main results are as follows:
\begin{itemize}
\item We confirm the validity of the different calibrations based on the strength of the CO bandhead in the $K$ band to estimate spectral types of RSGs. The accuracy of $\pm2$ spectral types estimated by \citetalias{davies07} seems a good estimate.
\item We confirm membership of five RSG candidates in the open cluster RSGC3, and find that their average $v_{{\rm LSR}}$ is compatible within the uncertainty of our method with the average $v_{{\rm LSR}}$ of Ste~2. Our average velocity for RSGC3,  $v_{{\rm rad}}=95\pm7\:{\rm km}\,{\rm s}^{-1}$, 
when compared to the rotation curve of the Galaxy in this direction, indicates that the cluster is located at $d\ga5$~kpc.
\item We find several RSGs in the immediate vicinity of RSGC3. Eight of them have colour excesses and $v_{{\rm LSR}}$ compatible with those of cluster members. Photometric criteria suggest an even higher number of RSGs in the surrounding area, with $\ga50$ supergiants between the clusters and the association.
\item Five of the RSGs are strongly concentrated and define a candidate cluster, which we call Alicante~7. Analysis of 2MASS photometry for this cluster suggests that it contains at least nine RSGs. Analysis of the run of extinction in the direction to Alicante~7 and RSGC3 shows the distribution of absorbing material to be very similar in both directions, and  indicates that both clusters are located at a distance $d\approx6$~kpc. The RSGs in both clusters have identical colours and luminosities, suggesting that they have the same age $\tau$=16--20$\,$Myr \citep{clark09}.
\end{itemize}

The data presented here strongly point at the existence of an extended massive association surrounding RSGC3. If higher-precision dynamical data can confirm the association of RSGC3 and Alicante~7, together with other objects around the clusters, this structure would contain $\ga$6$0\,000\:M_{\sun}$, and likely $\approx$10$0\,000\:M_{\sun}$.

Our results offer good prospects for a wide-area search using high-multiplexing spectrographs. The techniques explored in this work offer an effective way of selecting interesting candidates and adequate criteria for the calibration of both spectral type and luminosity, which can be easily automated. In a future paper, we will present the results of such a survey.

\begin{acknowledgements}
 
We thank the referee for many helpful suggestions.

Based on observations collected at the Centro Astron\'omico Hispano 
Alem\'an (CAHA), operated jointly by the Max-Planck Institut für 
Astronomie and the Instituto de Astrof\'{\i}sica de Andaluc\'{\i}a (CSIC).
This research is partially supported by the Spanish Ministerio de
Ciencia e Innovaci\'on (MICINN) under
grants AYA2010-21697-C05-05 and CSD2006-70.  JSC acknowledges support
from an RCUK fellowship.

 This publication makes use of data products from 
the Two Micron All Sky Survey, which is a joint project of the University of
Massachusetts and the Infrared Processing and Analysis
Center/California Institute of Technology, funded by the National
Aeronautics and Space Administration and the National Science
Foundation.  UKIDSS uses the
UKIRT Wide Field Camera (WFCAM; Casali et al. 2007) and a photometric
system described in \citet{hewett06}. The pipeline processing and
science archive are described in \citet{hambly08}. The DENIS project has been partly 
funded by the SCIENCE and the HCM plans of the European Commission
under grants CT920791 and CT940627. Jean Claude Renault from IAP was the Project manager.  Observations were  
carried out thanks to the contribution of numerous students and young 
scientists from all involved institutes, under the supervision of  P. Fouqu\'e, survey astronomer resident in Chile.  This research has made use of the Vizier and Aladin services developed at the Centre de Donn\'ees
Astronomiques de Strasbourg, France.

\end{acknowledgements}

{}

\clearpage

\begin{sidewaystable}
\caption{Summary of RSG candidates in Alicante~7, with their observed properties $^{~a}$. {\bf Top
    panel: } Stars in the main RSG clump, including the spectroscopically confirmed members. {\bf Bottom panel:  }Stars with similar colours and $Q$, but fainter magnitudes.\label{tab:props}}
\begin{tabular}{lcccccccccccccc} 
\hline
\hline
ID & Name & \multicolumn{4}{c}{2MASS} & \multicolumn{4}{c}{GLIMPSE   {\it
  Spitzer}} & \multicolumn{5}{c}{{\it MSX}} \\
   &  (in 2MASS) &   $J$ & $H$    & $K_{{\rm S}}$ &  $Q$ & Offset\tablefootmark{b} &4.5$\mu$m & 5.8$\mu$m  & 8.0$\mu$m  &  Offset\tablefootmark{b} & A   &   C  &   D   &  E \\
\hline
&&&&&&&&\\
A1(F1S09)& 18442053$-$0328446  & $9.28\pm0.02$  & 7.24$\pm$0.07  & 6.21$\pm$0.03  & 0.21 & $0\farcs21$ &$-$  &  5.38 & 5.45 &   $0\farcs6$  &  5.37& 5.33& 5.00 & $-$  \\
A2& 18442268$-$0331172  & $9.62\pm0.02$  & 7.22$\pm$0.04  & 6.08$\pm$0.02  & 0.35 & $0\farcs14$ & $-$  &  5.12 &  5.11 & $1\farcs3$ &  5.19 & 4.84 & 5.09 & $-$ \\
A3& 18442616$-$0335276  & $8.58\pm0.02$  & 6.19$\pm$0.05  & 5.07$\pm$0.02  & 0.37 & $0\farcs20$&  $-$ &   $-$ &  4.07 & $1\farcs3$ & 3.60 & 2.55 & 2.45 & 1.92\\
A4(F1S07)& 18442796$-$0329425  & $8.06\pm0.02$  & 5.91$\pm$0.04  & 4.85$\pm$0.02  & 0.25 & $-$ &  $-$ &  $-$ & $-$ & $1\farcs4$ & 3.43 & 2.46 & 2.35 & 1.83\\
A5(F1S06)& 18442945$-$0330024  & $8.45\pm0.03$  & 6.39$\pm$0.03  & 5.41$\pm$0.02  & 0.30 & $0\farcs18$ & $-$ & 4.57 & 4.48 &  $-$ &  $-$ & $-$  & $-$ & $-$\\
A6(F1S08)& 18443037$-$0328470  & $7.34\pm0.03$  & 5.20$\pm$0.04  & 4.10$\pm$0.29  & 0.16:& $-$ &  $-$ & $-$ & $-$ &  $1\farcs5$ &  4.30 & 3.14 & 3.03 & 2.23 \\
A7& 18443885$-$0326135  & $9.27\pm0.03$  & 7.12$\pm$0.03  & 6.02$\pm$0.03  & 0.18 & $0\farcs25$ & $-$ & 4.59 & 4.39 & $1\farcs3$ & 4.55 & 3.72 & 3.54 & $-$ \\
A8(F1S04)& 18443941$-$0330003  & $8.66\pm0.03$  & 6.76$\pm$0.04  & 5.82$\pm$0.02  & 0.19 & $0\farcs09$&  $-$ &  5.14 &  5.09 & $0\farcs9$ &  5.12  & 4.96 & 5.24 &  $-$\\
A9(F1S02)& 18444686$-$0331074  & $8.78\pm0.03$  & 6.71$\pm$0.02  & 5.76$\pm$0.02  & 0.38 & $0\farcs12$ &  $-$ & 5.00 & 4.95 & $1\farcs5$ & 5.02 & 4.87 & 4.88 &  $-$ \\
\hline
&&&&&&&&\\
A10&18443099$-$0334530 & $9.81\pm0.02$ & $7.77\pm0.04$ & $6.80\pm0.03$ & 0.29&  $0\farcs15$ & $-$ &  6.01 &  6.01 & $1\farcs8$ &  6.16 &  $-$ & $-$ & $-$\\
A11(F1S05)&18443100$-$0330499 & $9.87\pm0.02$ & $7.93\pm0.05$ & $6.97\pm0.03$ &0.21 &  $0\farcs09$ &  $-$  & 6.27 &  6.30 &   $2\farcs0$ &  6.33 &   $-$ &  $-$ & $-$\\
A12&18443749$-$0325148 & $10.00\pm0.02$& $8.09\pm0.04$ & $7.12\pm0.02$	&0.17 & $0\farcs24$ & 6.73 &  6.31 & 6.24 & $1\farcs8$ &  5.80 & 5.14 & 4.64&  2.73\\
A13&18444368$-$0329439 & $9.71\pm0.02$ & $7.93\pm0.07$ & $7.05\pm0.02$	&$0.19$ &
$0\farcs07$ &  6.63 &  6.32 & 6.36 & 0.6 &  6.43 &   $-$ &  $-$ &  $-$\\
 \hline
\end{tabular}
\newline\\
$^{~(a)}${ $JHK_{{\rm S}}$ magnitudes are from 2MASS, with mid-IR ($\sim4$--25~$\mu$m)
magnitudes from the Galactic plane surveys of GLIMPSE/{\it Spitzer}
\citep{benjamin} and the {\it Midcourse Source
  Experiment (MSX)}  \citep{egan}.} \\
$^{~(b)}${GLIMPSE and {\it MSX} counterparts have been searched within a $5\arcsec$ radius.}
\end{sidewaystable}


\begin{thebibliography}{}

\bibitem[Alexander et al.(2009)]{alexander09} Alexander, M.J.,
  Kobulnicky, H.A., Clemens, D.P., et al. 2009, AJ, 137, 4824

\bibitem[Alvarez \& Mennessier(1997)]{am97}Alvarez, R., \& Mennessier, M.-O. 1997, A\&A, 317, 761

\bibitem[Benjamin et al. (2003)]{benjamin} Benjamin, R.A., Churchwell, E., Babler, B.L., et al. 2003, PASP, 115, 953

\bibitem[Boschi et al.(2003)]{boschi} Boschi, F., Munari, U., Sordo, R., \& Marrese, P.M. 2003, in Corradi, R.L.M., et al. (eds.), ASP Conf. Series, vol. 303, 535

\bibitem[Cabrera-Lavers et al.(2005)]{cabrera05} Cabrera-Lavers, A., Garz\'on, F., \& Hammersley, P.L. 2005, A\&A, 433, 173

\bibitem[Carquillat et al.(1997)]{carquillat97} Carquillat, J.M., Jaschek, C., Jaschek, M., \& Ginestet, N. 1997, A\&AS, 123, 5

\bibitem[Cenarro et al.(2001)]{cenarro01} Cenarro, A.J., Cardiel, N.,
  Gorgas, J., et al. 2001, MNRAS, 326, 959

\bibitem[Cenarro et al.(2009)]{cenarro09} Cenarro, A.J., Cardiel, N.,
 Vazdekis, A., \& Gorgas, J. 2009, MNRAS, 396, 1895

\bibitem[Clark \& Porter(2004)]{cp04} Clark, J.S., \& Porter, J.M. 2004, A\&A, 427, 839

\bibitem[Clark et al.(2009a)]{clark09}
Clark, J. S., Negueruela, I., Davies, B., et al. 2009a, A\&A, 
498, 109

\bibitem[Clark et al.(2009b)]{simonw51}
Clark, J. S., Davies, B., Najarro, F., et al. 2009b, A\&A, 504, 429

\bibitem[Clark et al.(2010)]{clark10}
Clark, J. S., Ritchie, B.W., \& Negueruela, I. 2010, A\&A, 514, A87

\bibitem[Clark et al.(2011)]{clark11}
Clark, J. S., Ritchie, B.W., \& Thompson, M.A. 2011, Bulletin de la Societ\'e Royale des Sciences de Li\`ege, in press


\bibitem[Davies et al.(2007)]{davies07}
Davies, B., Figer, D.F., Kudritzki, R.-P., et al. 2007, ApJ, 671, 781 (D07)

\bibitem[Davies et al.(2008)]{davies08}
Davies, B., Figer, D.F., Law, C.J. et al. 2008, ApJ, 676, 1016 

\bibitem[D\'{\i}az et al.(1989)]{diaz89} D\'{\i}az, A.I., Terlevich, E., \& Terlevich R., 1989, MNRAS, 239, 325


\bibitem[Draper et al.(2000)]{draper} Draper, P.W., Taylor, M. , \&
  Allan, A. 2000, Starlink User Note 139.12, R.A.L.

\bibitem[Egan et al.(2001)]{egan}
Egan, M. P., Price, S. D., Gugliotti, G. M., 2001, BAAS, 34, 561

\bibitem[Figer et al.(2006)]{figer06}
Figer, D.F., MacKenty, J.W., Robberto, M., et al. 2006, ApJ, 643, 1166 

\bibitem[Garz\'on et al.(1997)]{garzon}
Garz\'on, F., L\'opez-Corredoira, M., Hammersley, P., et al. 1997, ApJ, 491, L31

\bibitem[Ginestet et al.(1994)]{ginestet94}
Ginestet, N., Carquillat, J.M., Jaschek, M., \& Jaschek, C. 1994, A\&AS, 108, 359 

\bibitem[Gredel et al.(2001)]{gredel01}
Gredel, R., Black, J.H., \& Yan, M. 2001,
A\&A, 375, 553

\bibitem[Hambly et al.(2008)]{hambly08} Hambly, N.C., Collins, R.S.,
  Cross, N.J.G., et al. 2008, MNRAS, 384, 637

\bibitem[Hewett et al.(2006)]{hewett06} Hewett, P.C., Warren, S.J.,
  Leggett, S.K., \& Hogkin, S.T. 2006, MNRAS, 367, 454

\bibitem[Hinkle et al.(2003)]{hinkle} Hinkle, K.H., Wallace, L., \& Livingston, W. 2003, BAAS, 35, 1260

\bibitem[Howarth et al.(1998)]{howarth}Howarth, I., Murray, J.,
  Mills, D., \& Berry, D.S. 1998, Starlink User Note 50.21, R.A.L.

\bibitem[Humphreys(1978)]{humphreys} Humphreys, R.M. 1978, ApJS, 38,
309

\bibitem[Humphreys \& McElroy(1984)]{hme84} Humphreys, R.M., \& McElroy, D.B. 1984, ApJ, 284, 565 

\bibitem[Keenan \& Heck(1994)]{kh94} 
Keenan, P.C. \& Heck, A. 1994, RMxAA, 29, 103

\bibitem[Keenan \& Pitts(1980)]{kp80} 
Keenan, P.C., \& Pitts, R.E. 1980, ApJS, 42, 541

\bibitem[Keenan \& McNeil(1989)]{keenan} 
Keenan, P.C., \& McNeil, R.C. 1989, ApJS, 71, 245

\bibitem[Kirkpatrick et al.(1991)]{kirkpatrick91} Kirkpatrick, J.D., Henry, T.J., \& McCarthy, D.W., Jr. 1991, ApJS 77, 417

\bibitem[Levesque et al.(2005)]{levesque} Levesque, E.M., Massey, P.,
  Olsen, K.A.G., et al. 2005, ApJ, 628, 973 

\bibitem[van Loon et al.(2005)]{vanloon05} van Loon, J.Th., Cioni, M.-R. L., Zijlstra, A.A., \& Loup, C. 2005, A\&A, 438, 273

\bibitem[Mallik(1997)]{mallik97} Mallik S.V., 1997,
A\&AS, 124, 359

\bibitem[Marco et al.(2007)]{marco} Marco, A., Negueruela, I., \& Motch, C. 2007, ASPC, 361, 388

\bibitem[Marigo et al.(2008)]{marigo}Marigo, P., Girardi, L.,
  Bressan, A., et al. 2008, A\&A, 482, 883

\bibitem[Marrese et al.(2003)]{marrese}Marrese, P.M., Boschi, F., \& Munari, U. 2003, A\&A, 406, 995


\bibitem[Mermilliod et al.(2008)]{mermilliod} Mermilliod, J.C., Mayor, M., \& Udry, S. 2008, A\&A, 485, 303

\bibitem[Meynet \& Maeder(2000)]{meynet00}
Meynet, G., \& Maeder, A. 2000, A\&A, 361, 101

\bibitem[Monet et al.(2003)]{monet03} Monet, D.G., Levine, S.E., Canzian, B., et al. 2003, AJ, 125, 984

\bibitem[Munari \& Tomasella(1999)]{munarit99} Munari, U., \& Tomasella, L. 1999, A\&AS, 137, 521

\bibitem[Munari et al.(2008)]{munari08} Munari, U., Tomasella, L., Fiorucci, M., et al. 2008, A\&A, 488, 969

\bibitem[Negueruela et al.(2010a)]{neg10} Negueruela, I., Gonz\'alez-Fern\'andez, C., Marco, A., Clark, J.S., \& Mart\'{\i}nez-N\'u\~{n}ez, S. 2010a, A\&A, 513, A74

\bibitem[Negueruela et al.(2010b)]{negwd1} Negueruela, I., Clark, J.S., \& Ritchie, B.W. 2010b, A\&A, 516, A78

\bibitem[Pandey et al.(2005)]{pandey} Pandey, A.K., Upadhyay, K., Ogura, K., et al. 2005, MNRAS, 358, 1290

\bibitem[Ramsey(1981)]{ramsey} Ramsey, L.W. 1981, AJ, 86, 557

\bibitem[Rathborne et al.(2009)]{rathborne09}
Rathborne, J.M., Johnson, A.M., Jackson, J. M., et al. 2009, ApJS, 182, 131

\bibitem[Rayner et al.(2009)]{rayner09} Rayner, J.T., Cushing, M.C., \& Vacca, W.D. 2009, ApJS, 185, 289

\bibitem[Shortridge et al.(1997)]{shortridge}Shortridge, K., Meyerdicks, H., 
Currie, M., et al. 1997, Starlink User Note 86.15, R.A.L.

\bibitem[Skrutskie et al.(2006)]{skru06} Skrutskie, M.F., Cutri, R.M.,
  Stiening, R. 2006, AJ, 131, 1163

\bibitem[Strai\v{z}ys \& Lazauskait\.{e}(2009)]{straizys09}
Strai\v{z}ys, V., \& Lazauskait\.{e}, R. 2009, Balt. Astr., 18, 19


\bibitem[Turner(1980)]{turner80}
Turner, D.G., 1980, ApJ, 240, 137



\bibitem[Zhou(1991)]{zhou91} Zhou, X., 1991, A\&A,
248, 367

\bibitem[Zhu et al.(1999)]{zhu99} Zhu, Z.X., Friedjun,g M., Zhao, G., et al. 1999, A\&AS, 140, 69

\end{thebibliography}
\end{document}